\documentclass[epsfig,showpacs]{revtex4}
\usepackage{epsfig,color}

\begin{document}
\title{Transition from participant to spectator fragmentation
       in Au+Au reaction between 60 AMeV and 150 AMeV}
\author{K. Zbiri${^1}$, A. Le F\`evre${^2}$, J. Aichelin${^1}$
, J.~{\L}ukasik$^2$, W. Reisdorf ${^2}$, F. Gulminelli ${^4}$ \\}
\author{the Aladin collaboration(U.~Lynen${^2}$,W.F.J.~M{\"u}ller${^2}$, H.~Orth${^2}$
C.~Schwarz${^2}$, C.~Sfienti${^2}$, W.~Trautmann${^2}$, K.~Turz{\'o}${^2}$,
B.~Zwiegli\'{n}ski${^3}$)\\}
\author{the INDRA collaboration
(
 J.L.~Charvet$^5$,
 A.~Chbihi$^3$,
 R.~Dayras$^6$,
 D.~Durand$^5$,
 J. D. Frankland$^3$,
 R.~Legrain$^5$,
 N.~Le Neindre$^6$,
 O.~Lopez$^4$,
 L.~Nalpas$^6$,c
 M.~Parlog$^{3,8}$,
 E.~Plagnol$^{9}$  
 M. F. Rivet$^6$,
 E.~Rosato$^{10}$,
 E.~Vient$^4$,
 M.~Vigilante$^{10}$,
  C.~Volant$^5$,
  J.P.~Wieleczko$^3$)\\} 

\address{
$^1$ SUBATECH, IN2P3-CNRS et Universit\'e, F-44072 Nantes Cedex 03, France.~\\
$^2$ Gesellschaft f{\"u}r Schwerionenforschung mbH, D-64291 Darmstadt,
Germany.~\\
$^3$ GANIL, CEA, IN2P3-CNRS, B.P.~5027, F-14021 Caen Cedex, France.~\\
$^4$ LPC, IN2P3-CNRS, ENSICAEN et Universit\'e, F-14050 Caen Cedex, France.~\\
$^5$ DAPNIA/SPhN, CEA/Saclay, 91191 Gif sur Yvette Cedex, France.~\\
$^6$ Institut de Physique Nucl\'eaire, IN2P3-CNRS, 91406 Orsay Cedex,
France.~\\
$^7$ IPN Lyon, IN2P3-CNRS et Universit\'e, F-69622 Villeurbanne Cedex, France.
\\
$^8$ Nuclear Institute for Physics and Nuclear Engineering, Bucharest,
Romania.~\\
$^9$ APC,F-75025 Paris Cedex 13,France \\
$^{10}$ Dipartimento di Scienze Fisiche, Univ.~di Napoli, I-180126 Napoli,
Italy.~}

\begin{abstract}
Using the quantum molecular dynamics approach, we analyze   
the results of the recent INDRA Au+Au experiments at GSI in the energy range between
60 AMeV and 150 AMeV. It turns out that in this energy region the transition toward
a participant-spectator scenario takes place. 
The large Au+Au system displays in the simulations as in the  
experiment simultaneously dynamical and statistical behavior which we analyze
in detail: The composition of fragments close to midrapidity follows statistical laws 
and the system shows bi-modality, i.e. a sudden transition between different 
fragmentation pattern as a function of the centrality as expected for a phase transition.
The fragment spectra at small and large rapidities, on the other hand, are determined by
dynamics and the system as a whole does not come to equilibrium, an observation which is 
confirmed by FOPI experiments for the same system.
\end{abstract}

\maketitle

\today
\section{INTRODUCTION}
Two decades after its discovery, 
 the rich phenomenology of multifragmentation has been widely explored 
(for recent references see, e.g., \cite{indra1,indra2}). It has been experimentally 
shown that in one single heavy ion collision many intermediate mass fragments (IMF's) 
are produced, where IMF's are defined as fragments with        
$3\le Z \le 25$. The upper limit is chosen to eliminate fission fragments.
Nevertheless, some of the key questions are still not answered.       
One of these, perhaps the most central one in order to come to a better 
understanding, is the question how fragments are formed. 
There are two reasons for this. First of all, under the
keyword multifragmentation two different processes are discussed which may be 
widely different in their physical origin. At low beam energies, the highest multiplicity   
of IMF's
is observed in central collisions. 
Fragments are formed from the matter in the geometrical overlap between projectile and target (participant matter). 
With increasing beam energy
the multiplicity of IMF's in central collisions decreases. At high beam energies, 
above several hundreds of AMeV, central collisions are that violent that 
only small nuclei, mainly up to mass $A=4$, survive and, therefore, the multiplicity of IMF's is low. 
Here, the largest IMF    
multiplicity is found in semi-peripheral reactions, and the fragments originate  
from those nucleons which are NOT in the geometrical overlap zone of projectile 
and target, the so-called spectator matter. In this case, particles from the    
interaction zone penetrate into the spectator matter and cause its disintegration 
into IMF's. The mean kinetic energy per nucleon of the fragments is lower than
in central collisions at low beam energy \cite{alad}.
It is all but clear whether the two processes, the one forming       
fragments from hot (energy per particle well above the binding energy 
of fragment nucleons \cite{alad}) and dense matter and the other forming fragments from
rather cold matter (energy per fragment below the  binding energy \cite{alad}) and at around normal nuclear     
matter density, have the same physical origin.    

Second, although completely different in their origin, 
statistical and dynamical models predict very similar results for several key 
observables. In the statistical or equilibrated source scenario \cite{fi,bon95,ber,QSM}
it is assumed that at a density which is a fraction of the normal nuclear matter density
the interaction among the constituents suddenly stops (freezes out) and 
that the relative fragment abundances at that moment are given by the phase 
space at the freeze out volume. Thus this model assumes that at the
latest at freeze out the system is in thermal or statistical equilibrium.
The phase space is calculated 
either in a microcanonical or in a grand canonical approach. 
In either case, it is assumed 
that, at the end,
the average thermal energy of the fragments is independent of the fragment size
and, neglecting Coulomb interaction and an eventual collective flow, equals 3/2
$T$ in a            
grand-canonical formulation. For energies larger than 50 AMeV the mass
yield of IMF's  follows a power law or an exponential function
which can hardly be distinguished due to the small range of IMF
masses.  

The dynamical approach presented in ref.\cite{HI} considers 
multifragmentation to be a fast process in which the
nucleons do not have the time to come to equilibrium, similar to the shattering of 
glass. There the distribution of splinters follows also a power law although
it is certainly not thermal. In the fragmentation process, the nucleons forming a fragment keep their momentum which they
have initially due to Fermi motion.    
As shown by Goldhaber \cite{gol} this fast fragmentation
yields as well a mass independent average energy of the fragments of the order of 15 MeV and a spectrum
similar to a thermal one. This means that single particle spectra  cannot
qualitatively distinguish between an already initially (due to the Fermi motion) present 
momentum distribution and a momentum distribution created by collisions in the 
expanding system. One can argue that the average energy should differ by a factor of two
(the average energy of 15 MeV due to Fermi motion as compared to a maximal thermal energy
(3/2 $T$) of 7.5 MeV because beyond a temperature of 5 MeV fragments are not stable anymore).  
If a transverse flow builds up during the expansion and in view of the additional 
Coulomb energy a distinction of the two slopes is only possible at high fragment kinetic energies. 
There the statistical error of the present experiments is too large for a 
distinction.

The scenario of a fast multifragmentation is also predicted by  
transport theories 
which describe the time evolution of the reaction starting from the
initially separated projectile and target nuclei
until the formation of the finally observed fragments. This models are either
based on true n-body approaches \cite{har1,aic,hartn,reg,AMD,amdind}
or on the Boltzmann Uehling Uhlenbeck approach
with fluctuating forces\cite{sur1}.
In the former approach fragments are to a large extent
initial correlations which have survived the heavy ion reaction. 
It is a challenge to understand why these initial-final state
correlations produce seemingly the same results as the statistical models. 
The systems which have
been investigated in these simulations so far are of moderate size.    
The recent Au+Au experiments of the INDRA collaboration at GSI in the beam energy range 
between 60 AMeV and 150 AMeV investigate really heavy systems (which may come closer
to equilibrium than lighter systems) with one of the most advanced
$4\pi$ detectors in the most interesting energy regime. In addition, the results can be compared   
with older experiments from the FOPI collaboration which cover a partially different phase space.
Therefore it is possible to cross check the results and to control the filters.  
Putting both experiments together a very detailed picture of the interaction should emerge.
This triggered a renewed effort to identify the origin of multifragmentation. For the 
experimental details of the INDRA experiment see ref. \cite{indra1}.

We start out in chapter two with an introduction to the Quantum Molecular 
Dynamics approach which we use to simulate the heavy ion reactions. 
Chapter 3 is devoted to the challenge to compare simulations with selected 
events. We will discuss in  detail how the detector acceptance changes the 
4$\pi$ particle distributions which are obtained in the simulation programs. 
There we discuss as well the importance to select the events in the same 
way as the experiments do. The much easier 
way to classify the theoretical simulation events according to 
the impact parameter picks events which can hardly be compared 
with an experimentally accessible event selection.
In chapter 4 we discuss
the global event structure and demonstrate that QMD produces well the
experimental centrality classes. Chapters 5 and 6 
are devoted to central
collisions. Chapter 5 presents a comparison of the theory with 
details of the reaction, like particle multiplicities. Chapter 6 presents the
new features obtained in the simulation: a) Midrapidity fragments    
are formed most probably in equal parts from projectile and target 
nucleons in contradiction to smaller systems \cite{reg} and b) the dynamical 
properties of the fragment source are strongly dependent on the fragment    
mass. Hence mixing of the nucleons in some regions of phase space 
occurs but a
kinetic equilibrium is not established. This is confirmed by the experiment.   
Chapter 7 is devoted to a study of the bi-modality in the QMD model.
Chapters 8 and 9 discuss in detail the reaction mechanism as seen in the
simulation. This study allows at the same time to identify the mechanism of
the fragment production and to observe how fragments survive the hot central
zone of the reaction. We see that in QMD fragments are surviving 
initial state correlations which have not been destroyed by binary
collisions. This mode of multifragmentation is similar to percolation with a   
percolation parameter above the critical value.
 Finally we will draw our conclusions.

\section{THE QMD MODEL}

The QMD model is a time dependent A-body theory to simulate the time
evolution of heavy ion reactions on an event by event basis. It is
based on a generalized variational principle. As every variational
approach it requires the choice of a test wave function $\Phi $.
In the QMD approach this is an A-body wave function with 6 A time
dependent parameters if the nuclear system contains A nucleons. 

To calculate the time evolution of the system we start out from the
action 

\[
S=\int _{t_{1}}^{t_{2}}{\mathcal{L}}[\Phi ,\Phi^{*}]dt\]

with the Lagrange functional\[
{\mathcal{L}}=<\Phi |i\hbar\frac{d}{dt}-H|\Phi >\]

The total time derivative includes the derivation with respect to
the parameters. The time evolution of the parameters is obtained by
the requirement that the action is stationary under the allowed variation
of the wave function. This leads to an Euler-Lagrange equation for
each time dependent parameter.

The basic assumption of the QMD model is that a test wave function
of the form

\[
\Phi ={\displaystyle \prod _{i=1}^{A_{c}+A_{p}}}\phi _{i}\]

with
\[
\phi _{i}(\overrightarrow{r},t)=(\frac{2}{L\pi })^{3/4}
e^{-(\overrightarrow{r}-\overrightarrow{r_{i}}(t))^{2}/4L}
e^{i(\overrightarrow{r}-\overrightarrow{r_{i}}(t))
\overrightarrow{p_{i}}(t)}e^{ip_{i}^{2}(t)t/2m}
\]
is a good approximation to the nuclear wave function. This means that
anti-symmetrization of the wave function \cite{AMD} is not essential at the energies
considered.

The time dependent parameters are $\overrightarrow{r_{i}}(t)$, $\overrightarrow{p_{i}}(t)$,
L is fixed and equals about 1.08 fm$^{2}$. 

Variation yields:

\[
\dot{\overrightarrow{r_{i}}}=\frac{\overrightarrow{p_{i}}}{m}+
\nabla _{\overrightarrow{p_{i}}}\sum _{j}<V_{ij}>=
\nabla _{\overrightarrow{p_{i}}}<H>\]

\[
\dot{\overrightarrow{p_{i}}}=-\nabla _{\overrightarrow{r_{i}}}
\sum _{j\neq i}<V_{ij}>=-\nabla _{\overrightarrow{r_{i}}}<H>\]

with
\[
<V_{ij}>=\int d^{3}rd^{3}r'\phi _{i}^{*}(\overrightarrow{r}')
\phi _{j}^{*}(\overrightarrow{r})V(\overrightarrow{r}',\overrightarrow{r})
\phi _{i}(\overrightarrow{r}')\phi _{j}(\overrightarrow{r}).
\]
These are the (i=1..N, N=$A_{P}+A_{T}$) time evolution equations which are 
solved numerically.
Thus the variational principle reduces the time evolution of the n-body
Schroedinger equation to the time evolution equations of 6
($A_{P}+A_{T}$) 
parameters to which a physical meaning can be attributed.

The nuclear dynamics of the QMD can also be translated into a semiclassical
scheme. The Wigner distribution function $f_{i}$ of the nucleon i
can be easily derived from the test wave functions (note that anti-symmetrization
is neglected):
\[
f_{i}(\overrightarrow{r},\overrightarrow{p},t)=\frac{1}{\pi ^{3}\hbar ^{3}}
e^{-(\overrightarrow{r}-\overrightarrow{r_{i}}(t))^{2}/2L}e^{(\overrightarrow{p}-\overrightarrow{p_{i}}(t))^{2}(2L/\hbar ^{2})}\]
and the total one body Wigner density is the sum of those of all nucleons.
The expectation value of the potential can be calculated with the help of the wave function or  
of the Wigner density. Hence the expectation value of the total Hamiltonian
reads
\[
<H>=<T>+<V>\]
where 
$<T>={\displaystyle \sum _{i}\frac{p_{i}^{2}}{2m_{i}}}$ 
and 
$<V>={\displaystyle \sum _{i}\sum _{j>i}\int
f_{i}(\overrightarrow{r},\overrightarrow{p},t)V^{ij}(\overrightarrow{r}',\overrightarrow{r})f_{j}(\overrightarrow{r}',
\overrightarrow{p'},t)d\overrightarrow{r}d\overrightarrow{r}'d\overrightarrow{p}d\overrightarrow{p}'}.$
The baryon-baryon potential $V_{ij}$ consists of Skyrme parametrization of the real part of the
Br\"uckner G-Matrix which is supplemented by an effective Coulomb interaction 
$V^{ij}_{Coul}$:
$V^{ij}=G^{ij}+V^{ij}_{Coul}$.
The former can be further subdivided into a part containing the contact Skyrme interaction 
and a contribution due to a finite range Yukawa-potential $V_{Yuk}^{ij}$. 
(In infinite matter the latter reduces
to a contact interaction as well but in finite nuclei it acts differently):
\begin{eqnarray*}
V^{ij}(\overrightarrow{r}',\overrightarrow{r})&=& 
V_{Skyrme}^{ij}+V_{Yuk}^{ij}+V_{Coul}^{ij} \\
&=&V_{ij}=t_{1}\delta (\overrightarrow{r}'-\overrightarrow{r})+
t_{2}\delta (\overrightarrow{r}'-\overrightarrow{r})\rho ^{\gamma -1}(\frac{\overrightarrow{r}'+\overrightarrow{r}}{2})+
t_{3}\frac{e^{\left\{ -|\overrightarrow{r}'-\overrightarrow{r}|/\mu \right\} }}{|\overrightarrow{r}'-\overrightarrow{r}|/\mu }
+\frac{Z_{i}Z_{j}e^{2}}{|\overrightarrow{r}'-\overrightarrow{r}|}
\end{eqnarray*}
The range of the Yukawa-potential is chosen as $ \mu = 1.5$ fm.   
$Z_{i},Z_{j}$ are the effective charges (
$\frac{Z_{proj}}{A_{proj}}$ for projectile nucleons, $\frac{Z_{targ}}{A_{targ}}$
for target nucleons) of the baryons i and j. The real part of the Bruckner G-matrix is density
dependent, which is reflected in the expression for $G^{ij}$.
The expectation value of G for the nucleon i is a function of the
interaction density $\rho _{int}^{i}$.
\[
\rho _{int}^{i}(\overrightarrow{r_{i}})={\displaystyle \sum _{j\neq i}}\int d^{3}rd^{3}r'\phi _{i}^{*}(\overrightarrow{r}')\phi _{j}^{*}(\overrightarrow{r})\delta (\overrightarrow{r}'-\overrightarrow{r})\phi _{i}(\overrightarrow{r}')\phi _{j}(\overrightarrow{r})\]

Note that the interaction density has twice the width of the single
particle density.

The imaginary part of the G-matrix acts like a collision term. In
the QMD simulations we restrict ourselves to binary collisions (two-body
level). The collisions are performed in a point-particle sense in
a similar way as in VUU or in cascade calculations: Two particles
may collide if they come closer than $r=\sqrt{\sigma /\pi }$ where
$\sigma $ is a parametrization of the free NN - cross section. A
collision does not take place if the final state phase space of the
scattered particles is already occupied by particles of the same kind
(Pauli blocking).

The initial values of the parameters are chosen in such a way that the
nucleons give proper densities and momentum distributions of the projectile
and target nuclei.
Fragments are determined here by a minimum spanning tree
procedure. At the end of the reaction all those nucleons are part
of a fragment which have a neighbour at a distance $r_{frag} \le 2.5 $fm.
$r_{frag}$ is a free parameter but it should not be smaller than
the force range in order that bound particles are counted as part of the
fragment. This radius is independent of the beam energy because 
in an expanding system two particles separate in coordinate space if they 
are not bound. Thus for each value of $r_{frag}$ one finds a time t
after which the minimum spanning tree procedure gives same fragment pattern 
as long as the system is expanding. This time t depends on energy. For the
simulations at 100 AMeV and 150 AMeV the fragment multiplicity has stabilized
before 200 fm/c.  
At 60 AMeV the relative velocities are small for this heavy system and 
it does not really expand. Therefore at 200 fm/c the fragments are 
not clearly separated in coordinate space. In this case the cluster 
distribution depends on  the value of $r_{frag}$. We have kept the standard 
value which gives a good overall description but
the results have to be treated with more caution.
The fragments have at 200 fm/c still some excitation energy. 

For further details of the QMD model we refer to ref.\cite{aic,hartn}. 
To compare the QMD simulations with experimental data as realistically   
as possible we built up a data base of about 100 000 QMD events over
a large impact parameter range. We have chosen a soft equation of       
state.
\section{Importance of the experimental filter for the comparison of experimental results
and QMD simulations}
A comparison between the results of the programs which simulate heavy ion reactions and the experiments
is all but easy. On the computer the positions and momenta of all particles are known at the end of the 
reaction. In experiments this is not the case even for the most advanced 4$\pi$ detectors. In peripheral reactions
the heavy residues disappear in the beam pipe or do not escape from target but even 
in the most central collisions the total charge of all measured fragments 
and light charged particles in a single event is not equal to the system charge 
but has instead a wide distribution. Particles hit the detector structure or their energy is below 
the detection threshold. In addition, the counters suffer from multiple hits which modify the 
particle identification. Therefore, theory and experiment can only be compared if one knows how the
detector would see a theoretical event. The software replica of the detector
which provides this information is called filter.
Its importance for a physical interpretation of the experimental results can hardly be overestimated.
For the experiments which we investigate here the filter which
takes into account the effects which are discussed above has been provided by the INDRA
collaboration \cite{filt}. 

If one is only interested in inclusive events, the filter serves only
to remove those particles which 
are not observed and 
to disentangle double hits in a given detector segment. For many physics questions, 
and they include multifragmentation, peripheral reactions are of very limited interest.
If one is interested in central events, it is difficult to 
underestimate the importance of a filter because it does not only correct the theoretical $4\pi$ simulation data 
for acceptance  but also determines the experimental centrality class to which the event belongs. 
\begin{figure}[hbt]
\epsfig{file=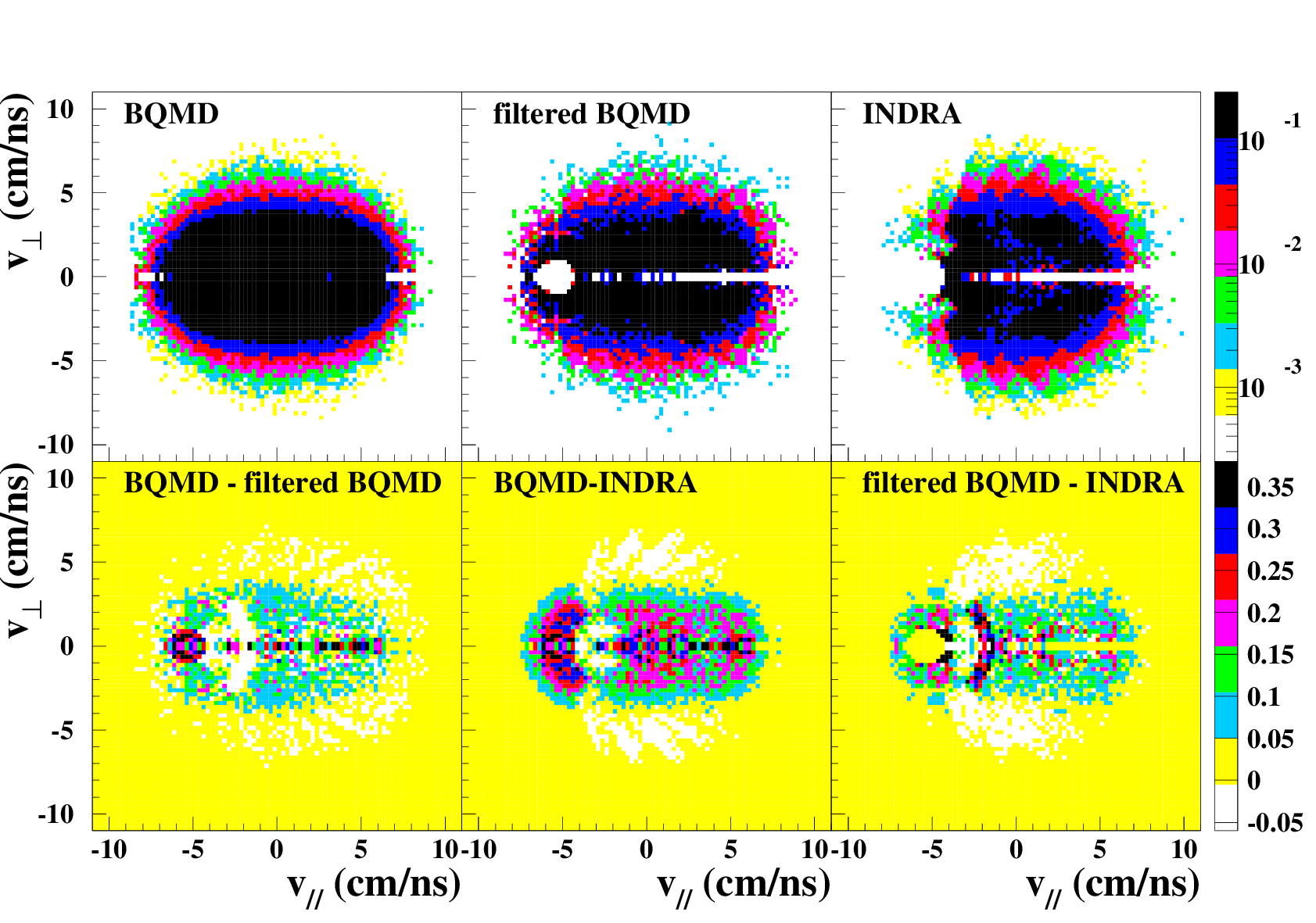,width=13cm} \vspace*{-.3cm}
\caption{Comparison of non filtered (top left) and filtered (top middle) QMD
distributions for Z=3 particles for the two most central bins with experimental
results (top right) for the reaction Au+Au at 60 AMeV. In order to show how the filter modifies the
events we display the difference between unfiltered and filtered QMD events
(bottom left), between unfiltered simulations and experiment (bottom
middle), as well as between filtered simulations and experiment (bottom
right).}
\label{a1}
\end{figure}  
The influence of the filter on fragment yields is usually much larger than
on light charged particles. 
How the filter modifies the raw simulation data on fragments is shown in fig.\ref{a1} where
we display the yield of central Au+Au reactions at 60 AMeV, the most critical energy,
in the transverse velocity / longitudinal velocity plane 
in the center of mass, for particles with Z=3. Only those events in which the total observed 
charge $Z_{tot}$ is larger than 78\% of the total charge of projectile and target 
are considered here. The top left (middle) figure shows
the simulation events before (after) we have applied the INDRA filter. Top right
one sees the experimental results.  The suppressed particles 
are displayed in the left bottom part of the plot. We see that the filter suppresses particles 
in the entire $v_{long}-v_{trans}$ plane.
On the first view this is 
astonishing because usually one expects that in forward direction the
large majority of particles is seen in the detectors. The suppression is strongest at small transverse
momenta. The difference between filtered events and data  
is displayed in the figure bottom right.  Although the filtered QMD events give a fragment distribution which
comes closer to the INDRA data than the unfiltered events (compare the middle and right figures in the bottom
row) the agreement is not at all perfect. We see that in the simulations there are 
too many fragments. The excess is concentrated along  an ellipse around midrapidity. This surplus 
appears at relatively high center of mass energies. This effect is especially pronounced for heavy 
fragments. There the filter creates fragments close to the beam velocity.
Therefore, if one averages over all events, the filtered simulation 
events show less stopping than the true events. In view of the above discussion this is due to
too little stopping in the simulations in this heavy system  subsequently amplified 
by the filter. This lack of stopping has not been observed for the
smaller Xe+Sn system at 50 AMeV \cite{reg}.   
The filter suppresses 
many more fragments with negative center of mass velocity ($v_{cm}$) than with positive 
$v_{cm}$. Hence  the filtered QMD events are no longer symmetric. Because, as we will see later,  
the simulation events produce very well the $E_{trans12}$ distribution ($E_{trans12}$ 
is the total transverse energy of all particles with charge Z= 1,2) and hence the
number of hard NN scatterings this lack of stopping is due to the mean field. Therefore 
the fragment pattern will not be influenced substantially.

\begin{figure}[h]
\epsfig{file=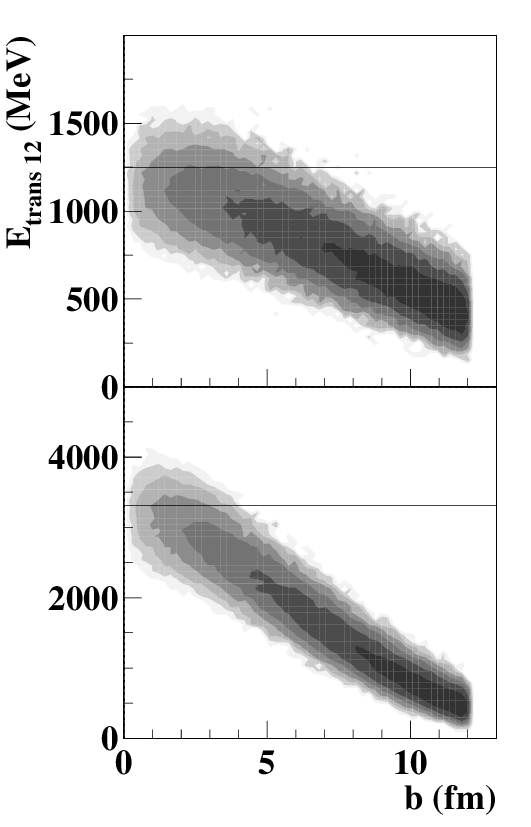,width=7cm} \vspace*{-.3cm}
\caption{The correlation between impact parameter and $E_{trans12}$ for Au+Au
reactions at 60 AMeV (top) and 150 AMeV (bottom) according to the QMD. The events above the   
cut in $E_{trans12}$ correspond
to a geometrical cross section of $\sigma =\pi $(1fm)$^2.$}
\label{a4}
\end{figure}

In the past, theoretical 
results for a given impact parameter have often been directly compared with data which   
have been selected according to their multiplicity or transverse energy 
\cite{amdind}. This may yield, as we show now, erroneous results.
In the experiment, the most central events are selected by requiring that
$E_{trans12} > 1246 (3313) $ MeV at 60 (150) AMeV which corresponds to a 
cross section of $\sigma =\pi $(1fm)$^2.$.
Because of the finite resolution in impact parameter, the so-defined event class is different from
the truly most central events with $b \le 1 $ fm which give the same cross section (fig.\ref{a4}).
\begin{figure}
\epsfig{file=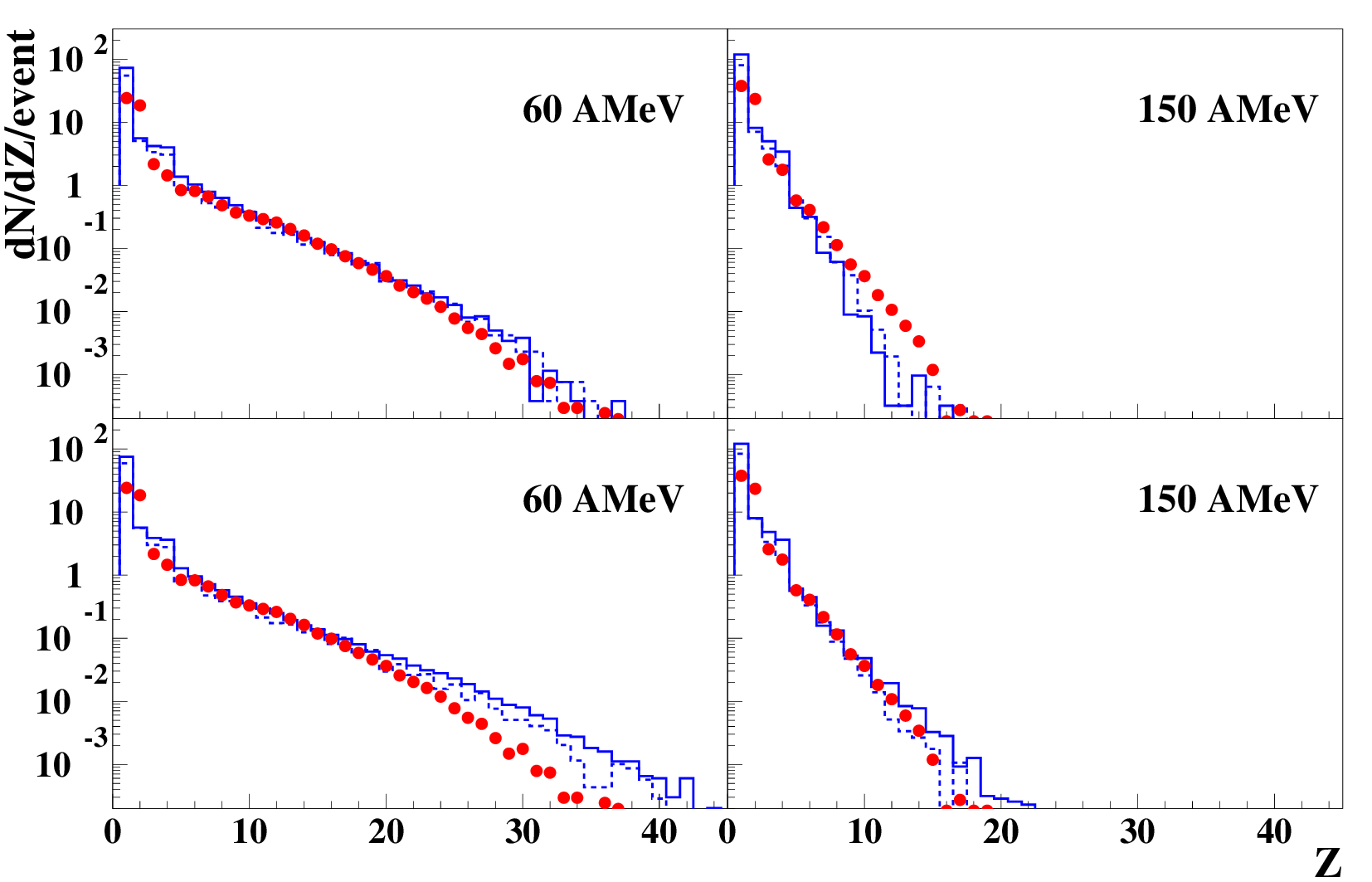,width=15cm} \vspace*{-.3cm}
\caption{The fragment yield of simulated events with $b \le 1$ fm (top) and $E_{trans12} 
\ge 1246 (3313)$ MeV (bottom) for Au+Au reactions 
as compared to the experimental data which have been selected by $E_{trans12}$.  Full (dashed) lines correspond to unfiltered 
(filtered) QMD events.}
\label{a4a}
\end{figure}
Therefore it is not astonishing that  physical quantities for the
two different choices of centrality differ as well. As an example we present in fig.
\ref{a4a} the fragment distributions for the reaction Au+Au at 60 AMeV
and 150 AMeV. With decreasing $E_{trans12}$ a smaller number of violent nucleon-nucleon
collisions have taken place and therefore heavier fragments can survive. We see
therefore a less steep fragment yield for the $E_{trans12}$
selection than for the impact parameter selection, especially for the
large fragments. The number of binary collisions  is inversely correlated with the
impact parameter and therefore also the probability that the initial-final
state correlations, which will be discussed in section IX, become destroyed.
The figure shows that quantitative comparisons require a cut in a variable which is
experimentally accessible. As we will see later, in the reaction at 60 AMeV, 
projectile and target form almost a compound system although in momentum space
the equilibration is not perfect. Consequently,
at the end of the calculation the nucleons remain very close in coordinate space.
This makes it very difficult to determine the fragments in the simulation events
and the systematical error is much larger than at higher energies where the fragments
are clearly separated in coordinate space at the end.

\section{global event structure}
\begin{figure}
\epsfig{file=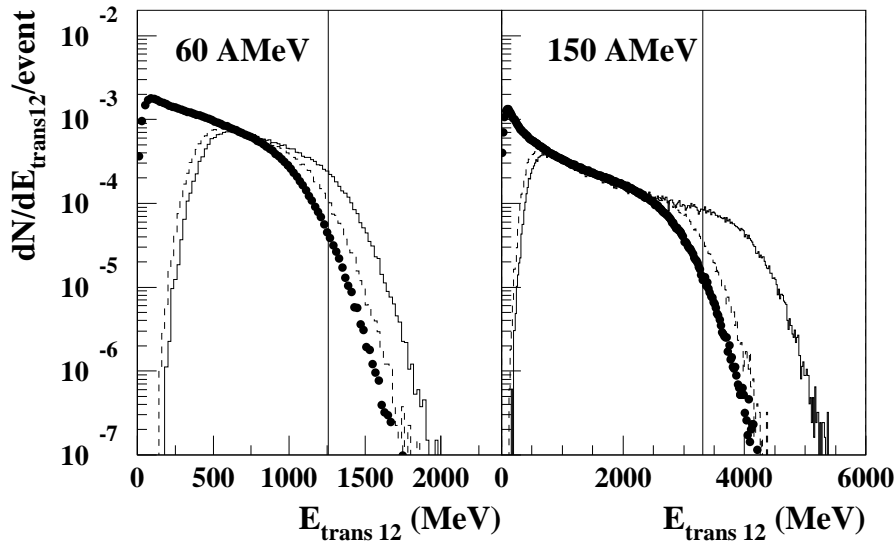,width=12cm} \vspace*{-.3cm}
\caption{The transverse energy distribution $dN/ E_{trans12}$ for Au+Au reactions 
at 60 AMeV (left) and 150 AMeV (right). The full line are the unfiltered
the dashed line the filtered QMD results. The QMD results are compared with the
data. The vertical lines show the experimental centrality cuts for
central events.}
\label{a44}
\end{figure}
In the analysis of the Au+Au reaction the energy of light particles
(Z= 1,2) ($E_{trans12}$) has been used for the event selection \cite{luk}. This differs from the
event selection criteria  
applied by the FOPI collaborations for the same reactions. We have, therefore,    
first of all, to check whether we can reproduce that quantity. If not,
it will not be meaningful to compare the simulations with the INDRA data  for    
selected centrality bins. In fig.\ref{a44} we display the 
($E_{trans12}$)  distributions, 
for unfiltered (full line) and 
filtered (dashed line) QMD events as well as for the INDRA data. The normalization is arbitrary because 
in our simulation the maximal  impact parameter is $b_{max}=12$ fm. The transverse
energy distribution of semi-peripheral and central collisions are well reproduced.
Please note that the $E_{trans12}$ distribution is also modified by the filter: 
The filter reduces $<E_{trans12}>$ by 13 (23)\% at 60 (150) AMeV.   

\section{Fragment distributions, multiplicities and spectra} 
\begin{figure}
\epsfig{file=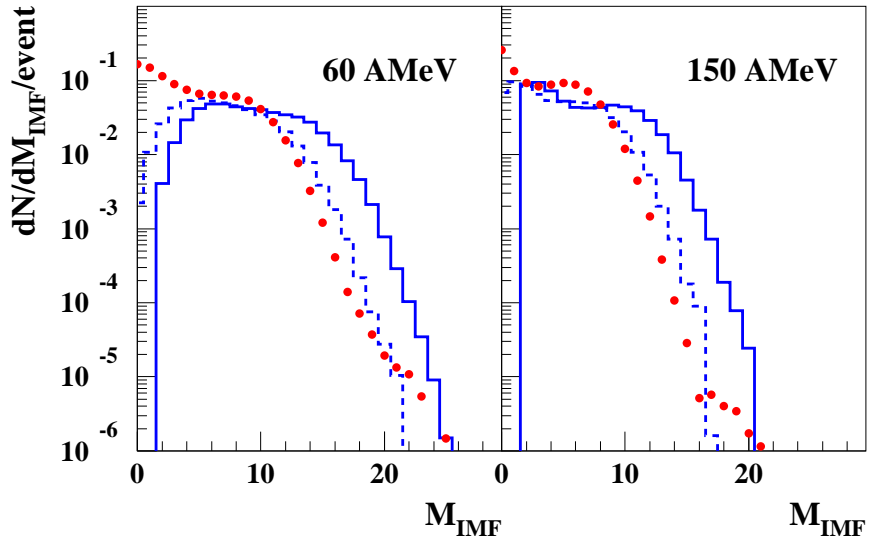,width=12cm} \vspace*{-.3cm}
\caption{Multiplicity distributions  of
intermediate mass fragments. We compare for all events unfiltered
(full line) and 
filtered (dashed line) QMD simulations with Au+Au data (60 AMeV left and 150 AMeV right).}
\label{a5}
\end{figure}

\begin{figure}
\epsfig{file=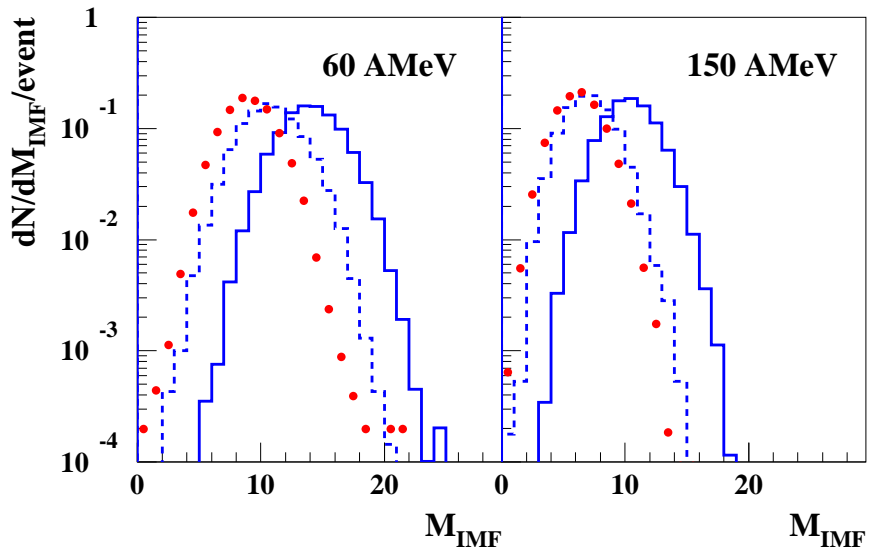,width=12cm} \vspace*{-.3cm}
\caption{Multiplicity distribution  of
intermediate mass fragments. We compare for central events unfiltered
(full line) and 
filtered (dashed line) QMD simulations with data (60 AMeV left and 150 AMeV right).}
\label{a6}
\end{figure}

\begin{figure}
\begin{minipage}{8cm}
\epsfig{file=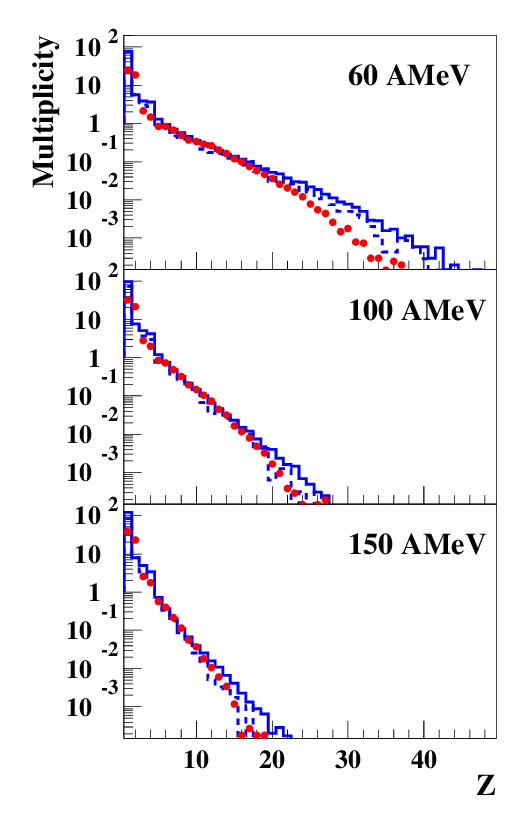,width=8.5cm} \vspace*{-.3cm}
\caption{Fragment distribution. We compare for central events
unfiltered (full line) and 
filtered QMD (dashed) simulations with data. 
}
\label{a7}
\end{minipage}
\begin{minipage}{8cm}
\epsfig{file=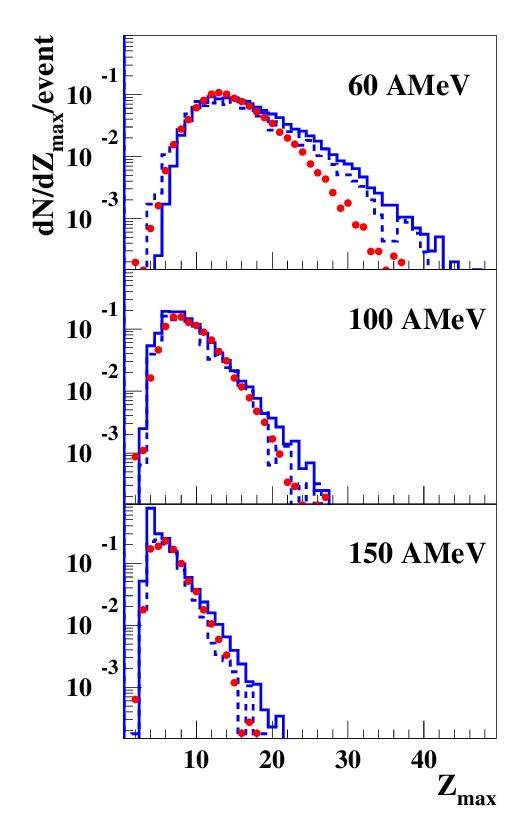,width=8.5cm} \vspace*{-.3cm}
\caption{Distribution of the heaviest fragment. We compare for central events
  unfiltered (full line) and 
filtered QMD (dashed) simulations with data.}
\label{a8}
\end{minipage}
\end{figure}

\begin{figure}
\epsfig{file=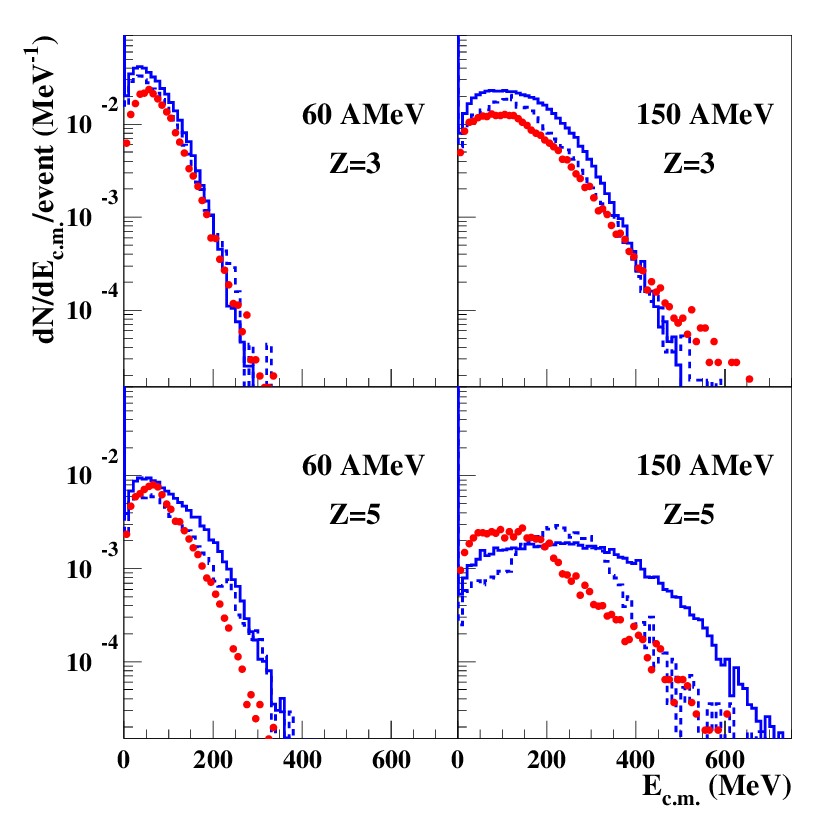,width=8.5cm} \vspace*{-.3cm}

\caption{Energy distribution of Z=3 and Z=5 fragments from    
central events, left for 60
AMeV, right for 150 AMeV.}
\label{fragp}
\end{figure}

After having seen that the transverse energy distribution of the light charged 
particles in the filtered simulations agrees well with that of the
INDRA data the next question is whether also fragments are reasonably reproduced.
In figs. \ref{a5} and \ref{a6} we display the multiplicity distribution of
intermediate mass fragments ($3\le  Z \le 25$). Fig.\ref{a5} displays 
the distribution for all impact parameters, fig.\ref{a6} shows the distribution for central 
events selected according to the experimental cut in $E_{trans12}$. These
central events correspond to a geometrical cross section of 3.14 fm$^2$.
We see, first of all, that the filter reduces the fragment multiplicity 
considerably and 
brings the distribution close to the experimental one. In fig.\ref{a5} one
sees the lack of events with a low multiplicity, a consequence of the fact that
we have stopped the simulations at b=12 fm. For central Au+Au reactions at 60
AMeV the filtered QMD events give the right form of the distribution
but over-predict slightly the multiplicity. At 150 AMeV
the form as well as the absolute value is well reproduced. Fig.\ref{a7}
shows the fragment yield and fig.\ref{a8} the charge of the heaviest fragment 
for central reactions in Au+Au at 60 AMeV (top), 100 AMeV (middle) and 150 AMeV (bottom).  
Again we see that these distributions are well described at 100 and 150 AMeV 
despite the considerable changes of this distribution between these two energies. At 60 AMeV
we see that we overpredict the fragment yield above Z= 25. This
points once more to the difficulty to identify the medium mass fragments in the
simulations at this energy due to their very small relative momentum.
The energy distribution for the Z=3 and Z=5 fragments is shown in fig.\ref{fragp}
for 60 AMeV and 150 AMeV. The slope is well reproduced by the QMD simulation 
in all cases but deviations occur at small $E_{cm}$ energies for Z=5 at 150 AMeV.
Experimentally the peak is close to the 
Coulomb barrier, whereas in the simulations the  fragments are less
stopped. This transparency is also seen for larger fragments.

\section{Is there an equilibrated source in central collisions?}
As said in the introduction, there are two different approaches to    
\begin{figure}
\epsfig{file=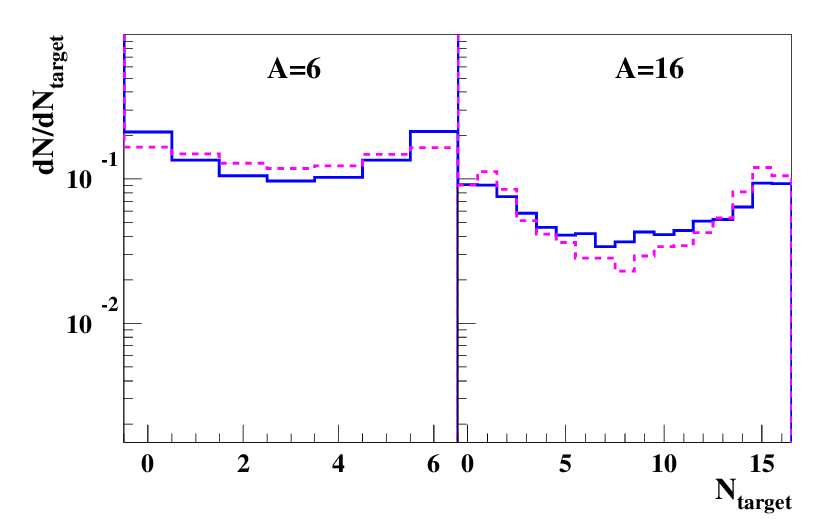,width=12cm} \vspace*{-.3cm}
\caption{QMD prediction for the composition of fragments in terms of nucleons which have been initially in the target
for A=6 (left) and and A=16 (right) in central Au+Au collisions ($b\le 3 \ $fm).
The full(dashed) line shows the distribution for 60 (150) AMeV.}
\label{a9}
\end{figure}
\begin{figure}
\epsfig{file=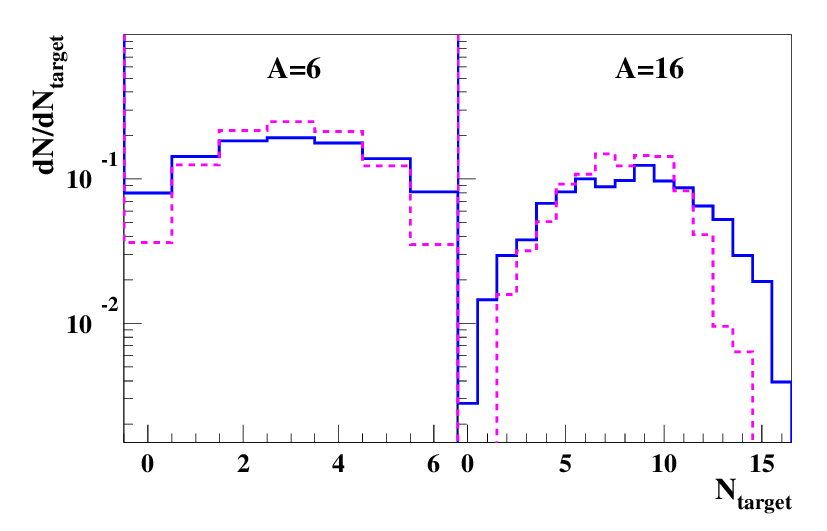,width=12cm} \vspace*{-.3cm}
\caption{QMD predictions for the composition of 
 mid-rapidity
fragments in terms of nucleons which have been initially in the target
for A=6 (left) 
 and A=16 (right) in central  Au+Au collisions 
($b\le 3 \ $ fm).
The full(dashed) line shows the distribution for 60 (150) AMeV.}
\label{a10}
\end{figure}
describe multifragmentation. If the statistical picture 
were correct, we would expect that in central collisions the nucleons 
in a fragment come in almost equal parts 
from projectile and target. For the system 50 AMeV Xe+Sn,
the dynamical
calculations showed that fragments are dominated either by projectile or 
by target
nucleons and only in rare cases fragments are formed in which both are present
with about the same weight \cite{goss}. 
\begin{figure}
\begin{center}
\epsfxsize=7.5cm
$$
\epsfbox{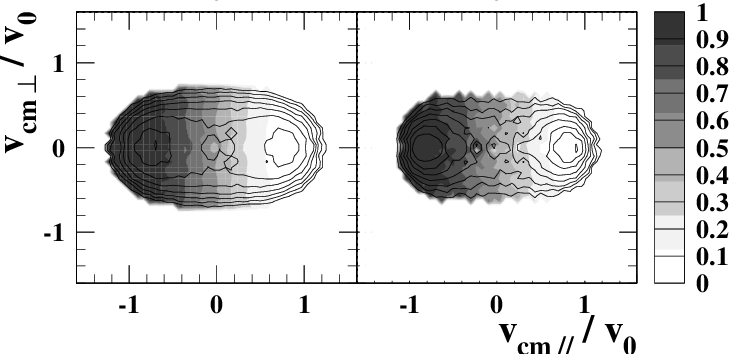}
$$
\end{center}
\caption{QMD predictions of the correlation between the fraction of  
entrained target nucleons and the longitudinal
fragment velocity in the center of mass for fragments with 
$ 6 \le Z \le 10$ in central Au+Au reactions}
\label{fra1}
\end{figure}
This situation is different for the heavier Au+Au system. Fig.\ref{a9} shows 
that in central collisions fragments of every mixture of projectile and 
target nucleons can be observed. Thus there exist fragments with the same number
of projectile and target nucleons. This is true for both energies and for 
different fragment sizes.
If we concentrate on fragments which are finally observed at
midrapidity ($60^\circ \le \theta_{cm} \le 120^\circ$), fig.\ref{a10}, 
we see this effect to be enhanced.     
\begin{figure}[htb]
\begin{center}
\epsfxsize=10.5cm
$$
\epsfbox{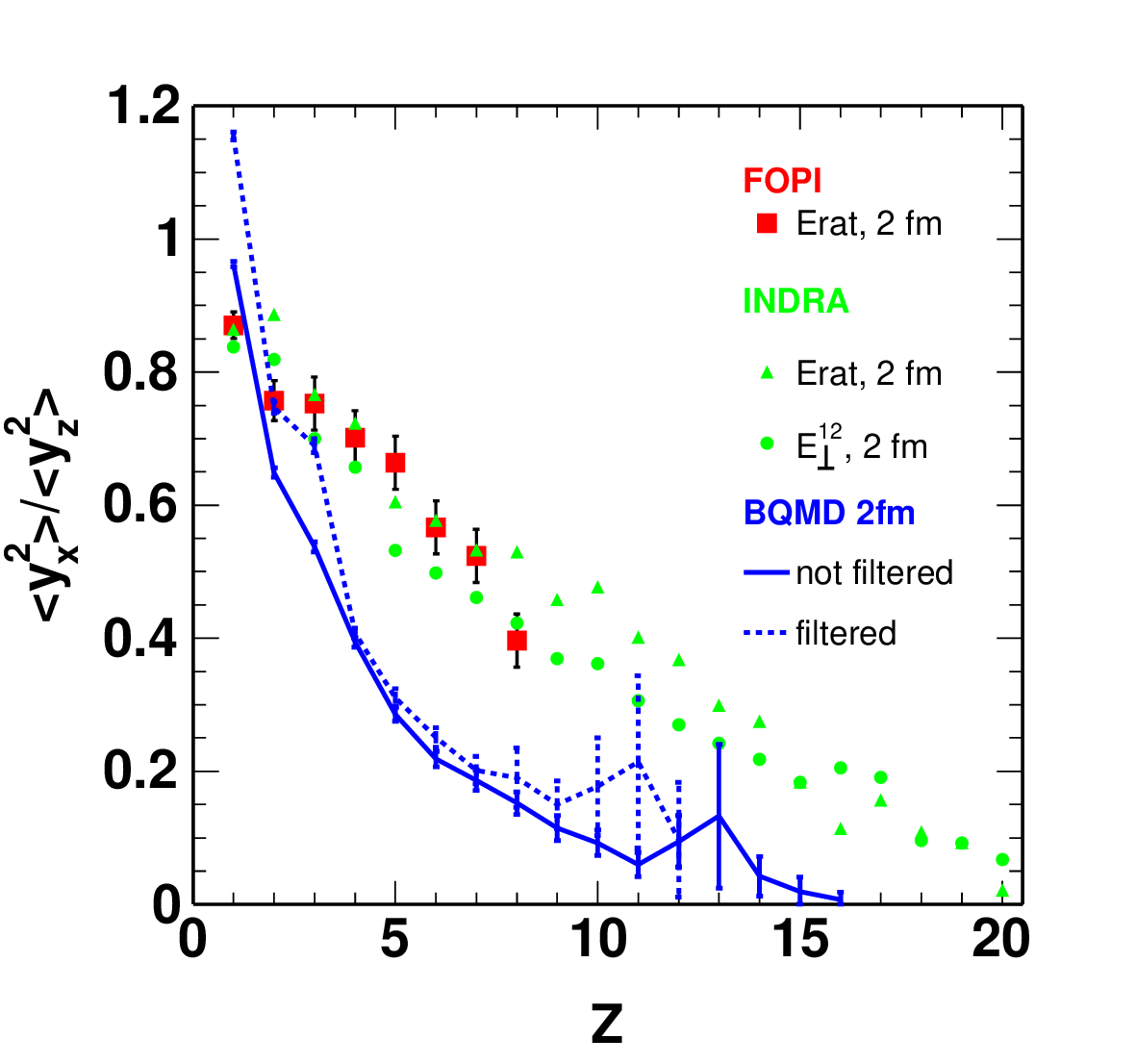}
$$
\caption{$R=\frac{<y^2x>}{<y^2z>}$ as a function of the fragment
charge for central events in Au +Au reactions at 150 AMeV measured by the 
INDRA and the FOPI collaboration \protect\cite{info}.               
The data are compared with the results of the QMD calculation.}
\label{a17}
\end{center}
\end{figure}
Here fragments
composed of a similar number of projectile and target nucleons dominate. Thus
central Au+Au collisions show complete mixing and therefore
statistical models can be employed to study  the fragment yields or 
fragment multiplicities.
Does this mean that the system has also reached equilibrium in the 
dynamical variables? To study this question
we investigate the correlation between the composition of     
a fragment and its velocity                                   
in the center of mass system. If equilibrium had been obtained we would see
a flat distribution because in equilibrium all nucleons have the same 
distribution independent of
their origin. The result of the QMD calculations, displayed in
fig.\ref{fra1}, shows, on the contrary, a strong 
correlation even at midrapidity where complete mixing has been
observed. Thus the dynamical degrees of freedom have not attained equilibrium.

To look further into this question  we calculate the mean squared rapidity 
variances in the impact
parameter direction, $<y^2_x>$, and in longitudinal direction,
$<y^2_z>$ for Au+Au at 150 AMeV. In a system in which the dynamical 
degrees of freedom
are equilibrated we expect a ratio 
$R= {<y^2_x>}/{<y^2_z>} = 1.$ QMD as well as both experiments, 
FOPI and INDRA, show that global
thermalization is not achieved. Fig.\ref{a17} displays the results. We see
that  protons are close to 
an equilibrium in the dynamical variables 
$(R\approx 0.9)$  but for fragments the value of R is well 
below 1. Even more, the ratio is a function of the fragment charge 
and decreases rapidly. In QMD this is understandable: As we will see 
the majority of fragments are surviving initial state     
correlations and the fragments are not decelerated substantially in           
longitudinal direction. Therefore their kinetic energy is, in first  
approximation, proportional to their atomic number A.                 
In terms of purely thermal models this means that the source properties 
depend on the fragments mass.
In order to make two independent 4$\pi$ experiments comparable \cite{info}, 
both, the ALADIN/INDRA as well as
the FOPI group, have made substantial efforts to determine the most 
central events. 
To determine central events one plots all events as a function of a 
certain centrality definition
($E_{rat}, E_{trans12}..$) and takes then those events with corresponds 
to a cross section $  \sigma < \pi \cdot 4 $ fm$^2$ assuming that the total 
reaction cross section is known. It is impossible to model
this criterion precisely in QMD. Therefore we have taken in the QMD events 
an impact parameter cut of $b = 2$ fm (see fig. \ref{a4a}).

\section{Bi-modality}
If a finite system undergoes a first order phase transition
bi-modality~\cite{Fran} is observed. It may even exist if a sizable
fraction of the initial momentum is not relaxed~\cite{fg} as it is the case in
heavy ion reactions. Bi-modality in systems with a phase transition 
means that for the same value of the control parameter the two phases, 
the ordered (liquid) and disordered (gas) one, are present.
Experimentally, the control parameter of the phase transition is very
difficult to access, if at all, and so one has to connect it to some
experimentally observable quantity. 

In order to study whether a liquid-gas phase transition can be observed 
in heavy ion reactions it has been suggested in ref.~\cite{tam} to study 
quasi-projectile  decay sorted by
quasi-target temperature, estimated from the total transverse energy of
light particles emitted at backward angles in the c.m. frame.
If a system is bi-modal in the same event class a liquid-like
phase (events with one large fragment) and a gas-like phase (events with
no large fragment) co-exist.

To quantify the bi-modality  one may define as in ref.~\cite{tam}  
\begin{equation}
a_2=(Z_{max}-Z_{max-1})/(Z_{max}+Z_{max-1})
\end{equation}  
where $Z_{max}$ is the charge of the largest fragment, while $Z_{max-1}$ is
the charge of the second largest fragment, both observed in the same event 
in the forward hemisphere.
If the system shows bi-modality we will observe in the same event
class two types of events: One with a large $a_2$ (one big fragment
with some very light ones) the other with small $a_2$ (two similarly sized   
fragments). Events with intermediate values of $a_2$ should be rare.

In the INDRA Au+Au experiments bi-modality has been indeed observed: 
In the same $E_{trans12}$ bin the distribution of the largest 
fragment shows two well separated maxima and $a_2$ as a function of         
$E_{trans12}$ varies very rapidly \cite{tam}. The question is whether this
observation can only be explained by a phase transition in a finite 
size system or whether alternative explanations
can be advanced.  

\begin{figure}[htb]
\begin{center}
\epsfxsize=10.5cm
$$
\epsfbox{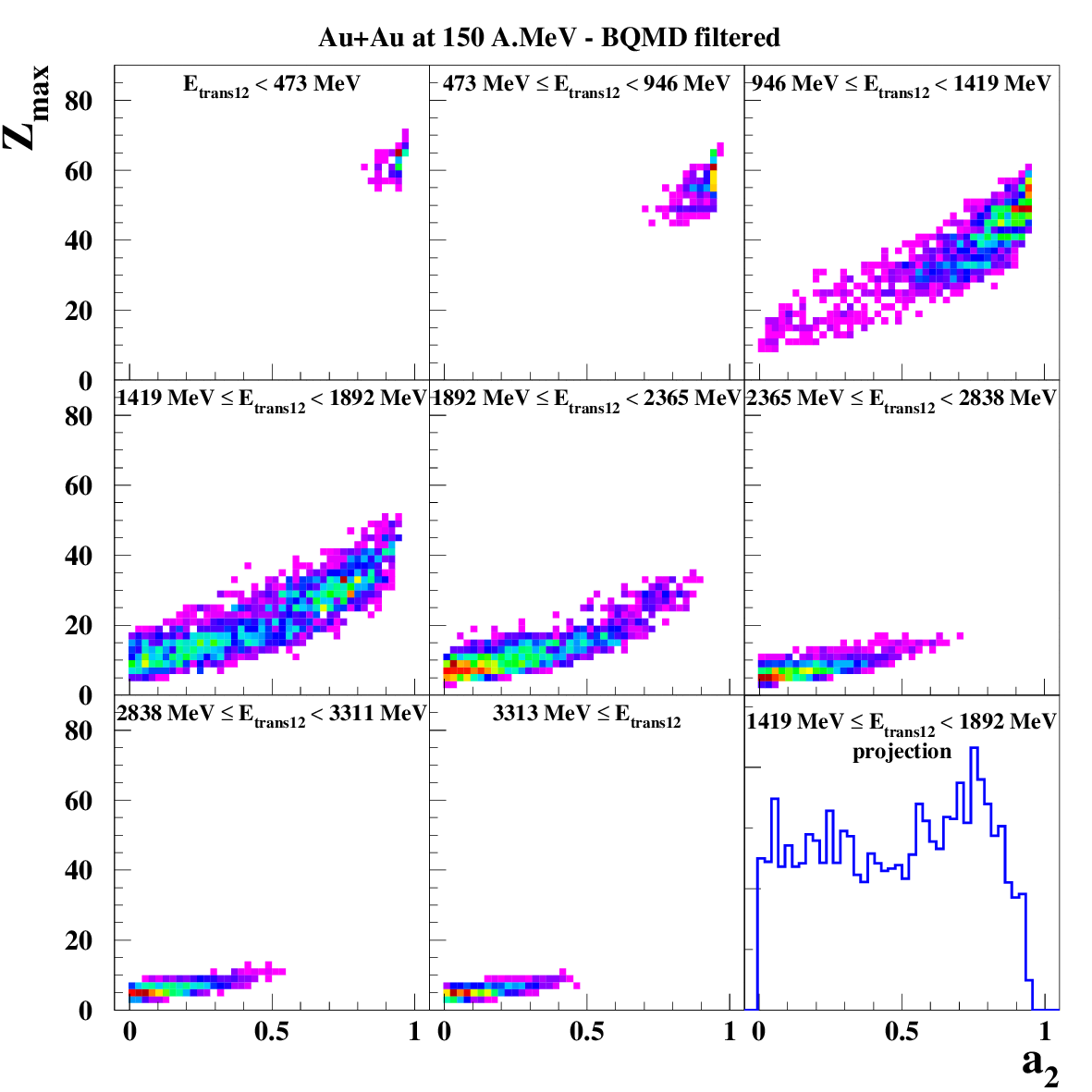}
$$
\caption{ $Z_{max}$ as a function of the asymmetry parameter for the
different experimental centrality bins for {\bf filtered} QMD events
at 150 AMeV.}  \label{bbb}
\end{center}
\end{figure}

As will be discussed in a separate paper, simulation programs like QMD show as
well bi-modality \cite{lza}. As an example we display in fig.~\ref{bbb}
$Z_{max}$ for filtered events at 150 AMeV. We made sure that also the 
unfiltered events have qualitatively the same structure. The bin
$1419 $ MeV $ \le E_{trans12} < 1892 $ MeV shows the presence
of two types of 
events with quite different $Z_{max}$ as may be inferred
from the right figure in the bottom row.
If one studies the origin of the bi-modality in QMD simulations one
realizes that at large impact parameters the momentum transfer
between projectile and target is not sufficient to decelerate the nuclei
substantially. At the end of the reaction we find two excited heavy remnants.   
At small impact parameters the stopping is not complete but the 
decelerated projectile and target remnants do not separate anymore. They 
remain connected by a bridge of matter from nucleons originating
from the overlap zone as will be discussed in section VIII. At the end of the
reaction this connected matter fragments. The break points are given by local
instabilities. Therefore small fragments of quite different sizes are formed.
This general behavior, that between projectile and target a bridge
of matter is formed in heavy ion collisions at intermediate impact parameters
has already been found in BUU calculations \cite{abuu}.

\begin{figure}[htb]
\begin{center}
\epsfxsize=10.5cm
$$
\epsfbox{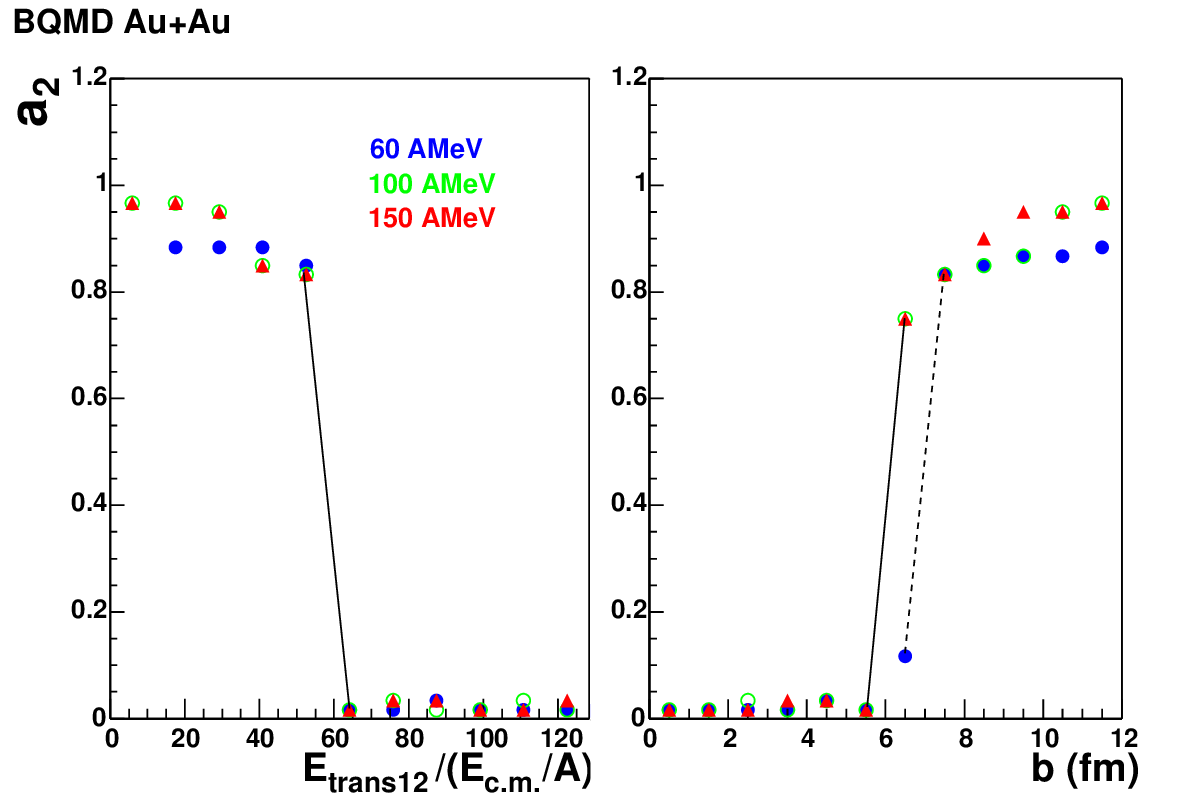}
$$
\caption{The most probable $a_2$ in unfiltered QMD simulations as a function of 
$E_{trans12}/E_{cm}$ 
(left) and as a function of the impact parameter (right). As in experiment we 
see in both cases a fast transition.}
\label{c5}
\end{center}
\end{figure}

The transition between the two reaction scenarios is rather sharp. Therefore we
see a sudden increase of the $a_2$ value if we increase the impact parameter
(fig. \ref{c5} right). Because the stopping and the impact parameter are 
strongly correlated we observe a similar increase if we plot $a_2$ as a 
function of $E_{trans12}$ as shown in fig. \ref{c5} left. Due to the increase 
of the nucleon-nucleon cross section with energy for a given impact parameter
the momentum transfer depends on the beam energy . Therefore the value of b 
for which this transition takes place varies with the beam energy. 
On the contrary,
the value of $E_{trans12}$, which measures directly the energy transfer, remains
constant as observed also in the INDRA experiments.

\section{The Dynamics of the Reaction}
\subsection{Transition between participant and spectator fragmentation}
\begin{figure}[htb]
\epsfxsize=12.0cm
$$
\epsfbox{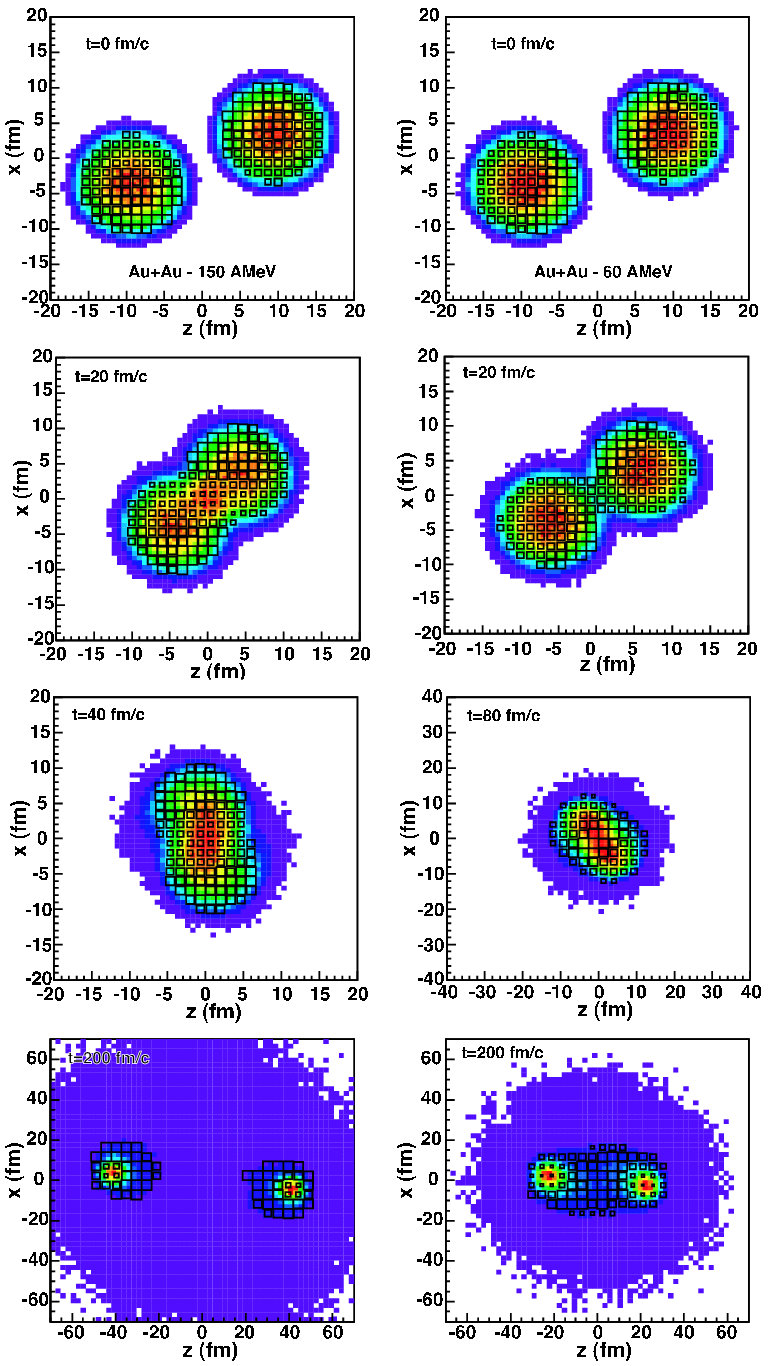}
$$
\caption{QMD predictions of the distribution of all nucleons (colors) and 
those nucleons which belong
finally to a fragment $\textrm{10}\leq A\leq \textrm{20}$ (boxes) for
collisions $ 6 \ge b \ge $8 fm for Au + Au at 150 AMeV (left) and
60 AMeV (right) from t= 0 (top) to t=200 fm/c(bottom)}
\label{a86}
\end{figure}

\begin{figure}[htb]
\epsfxsize=12.0cm
$$
\epsfbox{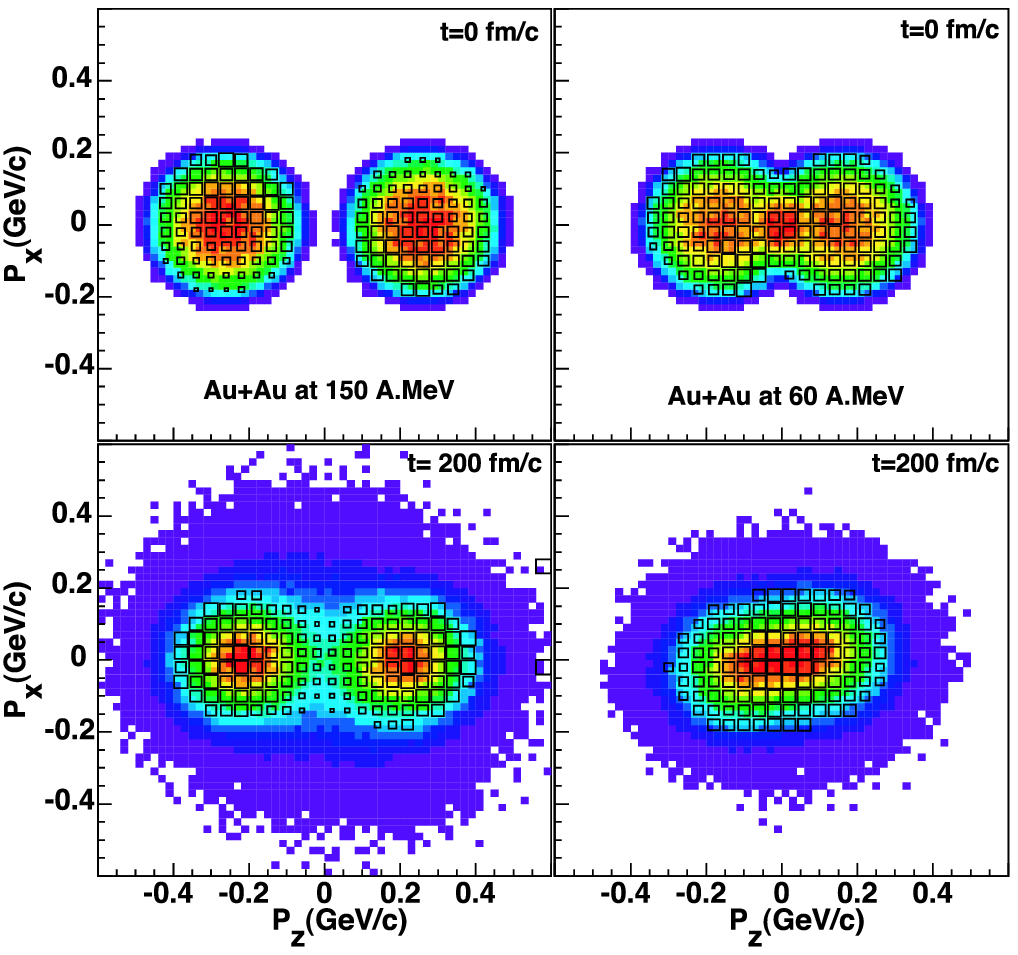}
$$
\caption{QMD prediction of the momentum space distribution of all 
nucleons (colors) and those nucleons which belong
finally to a fragment $\textrm{10}\leq A\leq \textrm{20}$ (boxes) 
for collisions
$ 6 \ge b \ge $8 fm for Au + Au at 150 AMeV (left) and
60 AMeV (right) at t= 0 (top) and t=200 fm/c(bottom)}
\label{a87}
\end{figure}

To study the evolution of the reaction mechanism from participant to spectator
dominated multifragmentation we use semicentral reactions 
$ 6\le b\le 8 $ fm and medium mass fragments ($ 10 \le A \le 20$). 
For this purpose we use now the fact that in the QMD simulations one knows 
the position and momentum of all particles at any given time point and 
therefore it is possible to study the history of those
nucleons which are finally part of the different fragments.
The color coding shows where the nucleons are, independent of whether 
these nucleons will finally be part of a fragment. The size of the 
squares gives the percentage of the nucleons (as compared to all nucleons) 
which end up finally in fragments of the selected
class (here $ 10 \le A \le 20$). We plot these both distributions for
different time steps, on the right hand side for the reaction at 60 AMeV and on
the left hand side for 150 AMeV in fig.
\ref{a86}. This figure is supplemented by fig.~\ref{a87} which shows the 
initial and final momentum distribution using the same
coding. We see, first of all, that at both energies the initial distribution 
of those nucleons which end finally up in fragments  is different  
from that of all nucleons. This means that strong initial-final state 
correlations are present which we study now in detail.
In coordinate space the nucleons which form finally fragments 
$10\le A\le 20$ are located toward the reaction partner. At 150 AMeV 
one sees clearly that they come from the spectator matter.
The time evolution for both energies is rather different and best seen 
if one compares the positions at 80fm/c of 60 AMeV with those at 40fm/c 
of 150 AMeV. At 60 AMeV we observe neck formation as at lower energies 
and the future fragment nucleons are concentrated in the neck, i.e. 
in the center of the reaction.
At 150 AMeV the nucleons show a completely different behavior. 
The future fragment nucleons are those which are not in the geometrical 
overlap of projectile and target. This is a clear indication
that between 60 AMeV and 150 AMeV the transition between participant 
and spectator fragmentation takes place, a transition which was 
believed to take place at considerably higher energies  and has
been observed at energies above 400 AMeV \cite{Schuet}.

In addition to the initial-final state correlations in coordinate space, 
there are also similar correlations in momentum space. At 150 AMeV 
future fragment nucleons have a transverse momentum away from the 
reaction zone (and thus the observed transverse fragment velocity is 
partially due to the selection of the fragment nucleons \cite{neba}). 
At 60 AMeV the correlation are less important but nucleons with a smaller
longitudinal momentum have a higher chance to be part of a IMF than 
those with a larger longitudinal momentum. 

\begin{figure}[htb]
\epsfxsize=12.0cm
$$
\epsfbox{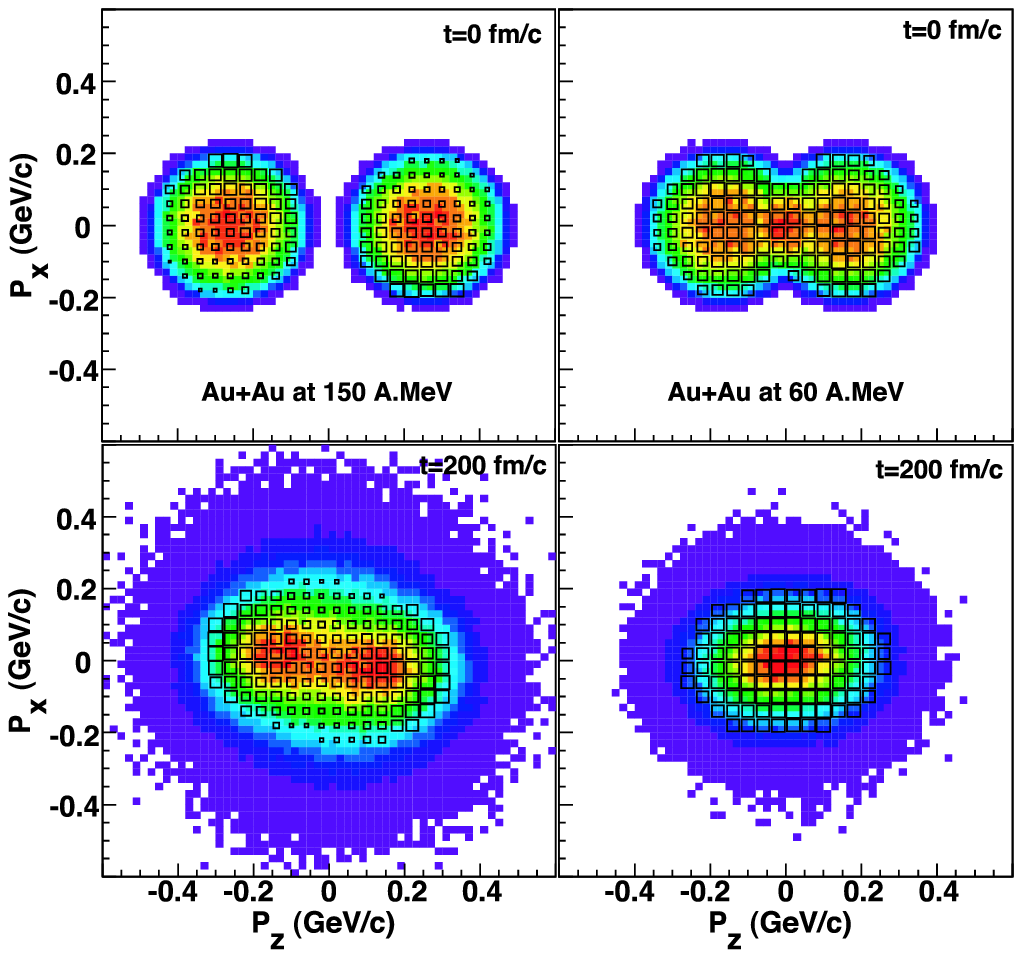}
$$
\caption{Momentum space distribution of all nucleons (colors) and those 
nucleons which belong
finally to a fragment $1\textrm{0}\leq A\leq \textrm{20}$ (boxes) for
central collisions $ 0 \ge b \ge 4 $ fm for Au + Au at 150 AMeV (left) and
60 AMeV (right) at t= 0 (top) and t=200 fm/c (bottom).}
\label{a99}
\end{figure}
Central collisions are rather similar to the semicentral ones at both 
energies. Again the fragment nucleons come dominantly from the overlap 
zone at 60 AMeV and from the spectator matter at 150 AMeV.
Therefore we show only the momentum space distributions
which are displayed in fig.\ref{a99}.
The average deceleration is of course much stronger as compared to 
the semicentral reactions but the fragments at 150 AMeV, coming from 
the spectator matter, are less influenced by this. 
They have still a quite large momentum. However, the matter at 
midrapidity is now that dense that some fragments are created forming 
the midrapidity source discussed above. The in-plane 
flow seen in semicentral collisions has almost disappeared, as expected.  
\begin{figure}[htb]
\epsfxsize=6.0cm
$$
\epsfbox{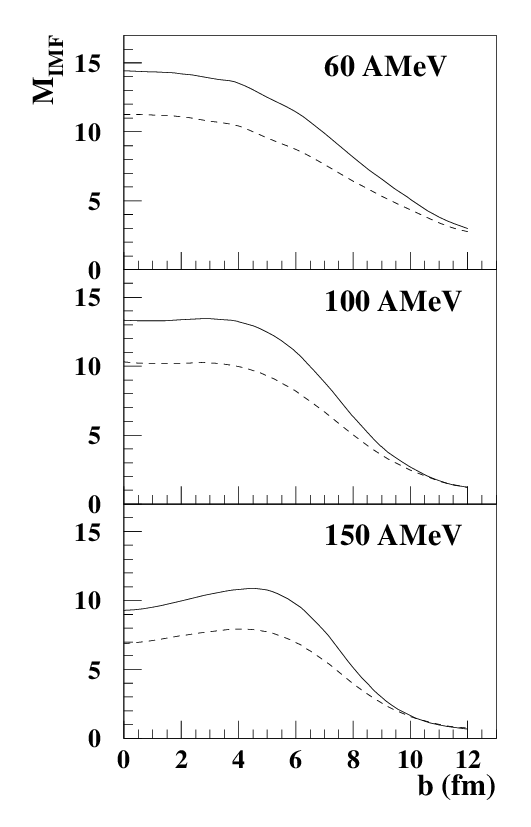}
$$
\caption{Multiplicity of intermediate mass fragments as a function of the impact parameter
for Au+Au reactions at 60 AMeV, 100 AMeV and 150 AMeV. The full (dashed) line shows unfiltered (filtered)
QMD events.}
\label{a88}
\end{figure}
This transition between participant and spectator fragmentation is 
also visible if one plots the multiplicity of IMF's as a function of 
the impact parameter, as done in fig. \ref{a88}. Whereas
at 60 AMeV the multiplicity peaks at $b=0$, at 150 AMeV semicentral 
events show a higher multiplicity. The filter modifies this observation 
which agrees with the data \cite{tsang,luka} only 
slightly. 

\subsection{Small IMF's come from many sources}  
\begin{figure}[htb]
\epsfxsize=12.0cm
$$
\epsfbox{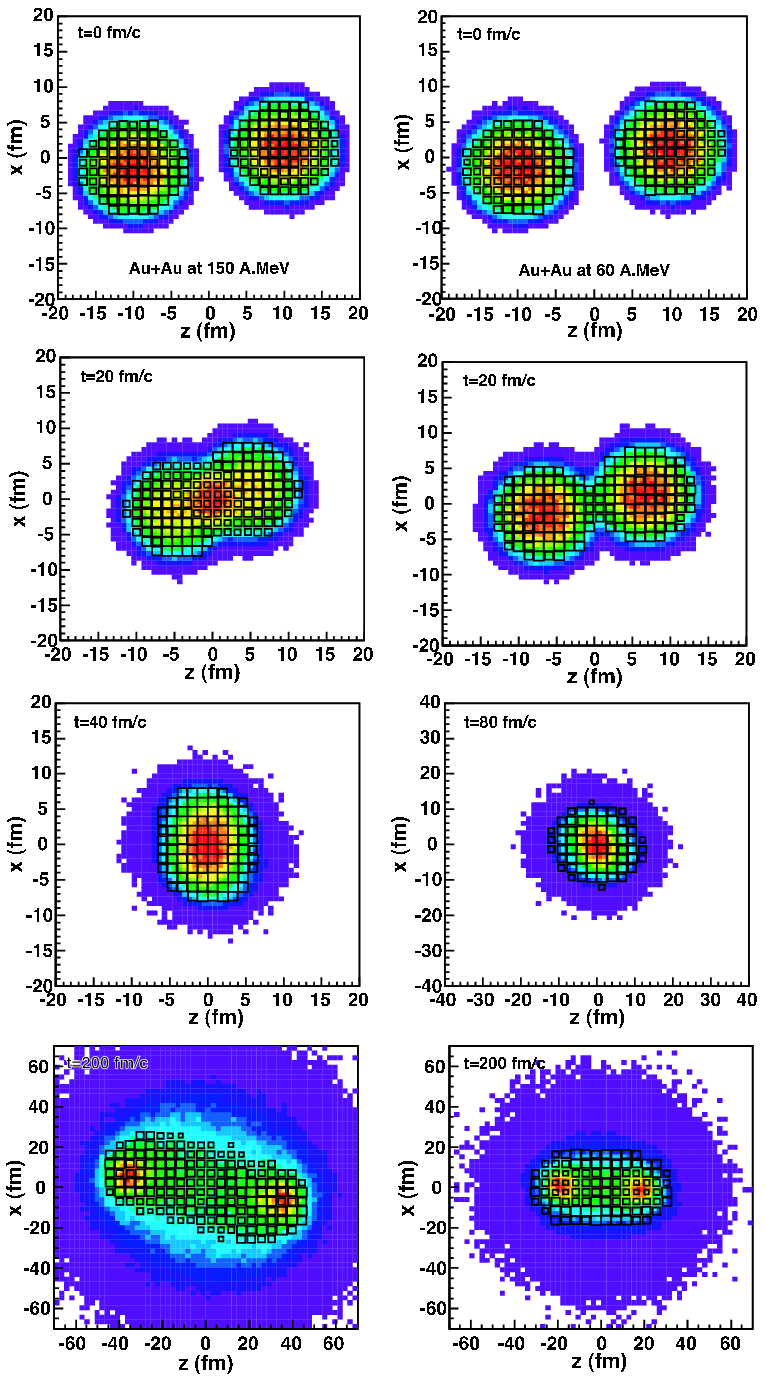}
$$
\caption{Distribution of all nucleons (colors) and those 
nucleons which belong
finally to a fragment $\textrm{6}\leq A\leq \textrm{10}$ (boxes) for
central collisions $ 0 \ge b \ge 4 $ fm for Au + Au at 150 AMeV (left) and
60 AMeV (right) 
 from t= 0 (top) to t=200 fm/c (bottom)}
\label{a55}
\end{figure}

\begin{figure}[htb]
\epsfxsize=12.0cm
$$
\epsfbox{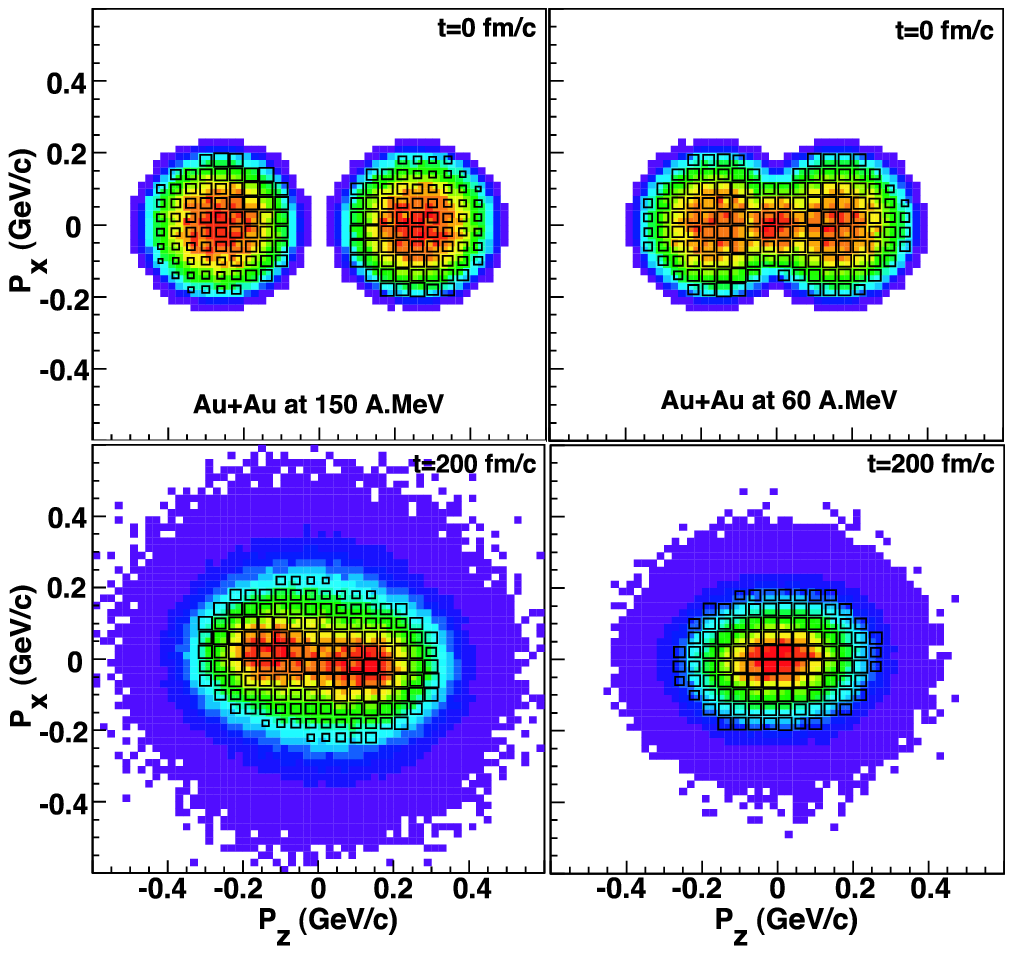}
$$
\caption{Momentum space distribution of all nucleons (colors) and those 
nucleons which belong
finally to a fragment $\textrm{6}\leq A\leq \textrm{10}$ (boxes) for
central collisions $ 0 \ge b \ge 4 $ fm for Au + Au at 150 AMeV (left) and
60 AMeV (right) at t= 0 (top) and t=200 fm/c(bottom)}
\label{a11}
\end{figure}

For semicentral collisions the initial-final correlations of those nucleons 
which end up in $1\textrm{0}\leq A\leq \textrm{20}$ fragments are almost identical
to that of $\textrm{6}\leq A\leq \textrm{10}$ fragments. Therefore we do not display
them. A difference can be observed in central collisions. The time
evolution in coordinate space is presented in 
fig.\ref{a55} whereas that in 
momentum space is presented in fig.\ref{a11}.
Here, at 150 AMeV, in addition to the fragments from the spectator matter
a midrapidity source develops (seen clearly in the second row of fig.\ref{a55}) 
which finally creates a bridge between target and projectile 
spectator fragments, seen in the bottom row, similar to what we have seen
for large fragments at 60 AMeV. This is reflected, of course, 
in momentum space (fig.\ref{a11}) where we see - in contradistinction to the 
$1\textrm{0}\leq A\leq \textrm{20}$ data - a midrapidity source.
It has never been observed before in
simulations of smaller systems that this midrapidity source, which reminds on the neck formation 
at lower energies, emits fragments of this size.  Also the initial-final state
correlations in coordinate space are much weaker than for the larger
fragment class, $1\textrm{0}\leq A\leq \textrm{20}$. We see that quite
a few of these fragments come from the participant matter.

\section{How fragments can survive the high density zone}
In section VI we have seen that part of the fragments are made of nucleons 
which have traversed the reaction zone. How this can happen we have
shown in ref. \cite{goss} 
for the system Xe+Sn.
There we have found
that fragments are made of nucleons which have passed the reaction zone
without having had collisions with a large transverse momentum transfer.
Nucleons with similar momenta which are close in coordinate space
suffer the potential interaction in the same way and therefore are 
collectively deviated by potential gradients. Therefore they leave 
collectively the interaction zone without the initial correlation among them
 being destroyed.  Therefore in QMD calculations fragments at forward
and backward rapidity present those initial state correlations 
which have not been destroyed by hard binary collisions. If these collisions
become more frequent, for example by increasing the beam energy and by
this reducing the influence of the Pauli blocking, less fragments are
observed, because more of the initial state correlations have been destroyed.

Here we explore whether this mechanism remains valid even for the large 
Au+Au system.
To study this question we make use of the fact that we know the
position and momentum of all nucleons during the whole reaction. This allows to
trace back the history of all nucleons and especially of those which
are finally part of a fragment i which contains A nucleons. 
For these nucleons we define three quantities, the average momentum
\begin{equation}
\vec P^i(t)=\frac{1}{A} {\displaystyle \sum _{j=1}^{A}}\vec p_j(t),
\end{equation}
the position of the center of mass of the fragment
\begin{equation}
\vec R^i(t)=\frac{1}{A} {\displaystyle \sum _{j=1}^{A}}\vec r_j(t),
\end{equation}
and the average transverse (with respect to the beam momentum) 
kinetic energy of the fragment nucleons in the fragment rest system
\begin{equation}
\Delta^{i}(t)={\displaystyle \sum _{j=1}^{A}}
\frac{(p_j^\perp(t)-P^{i\perp}(t))^{2}}{2m{A}}.
\end{equation}
 $\Delta^{i}(t)$ is sometimes sloppily called "fragment temperature".  
The finite value at t = 0 is due to the Fermi motion. We compare now
this ''fragment temperature'' with the environment.                       
The environment is defined by those nucleons which are at the same time 
closer than 2.5 fm to
the center of the fragment $\vec R^i(t)$ and also are not belonging to 
the fragment i. For those
nucleons we define as well the average kinetic energy
\begin{equation}
\Delta^{env}_{i}(t)={\displaystyle \sum _{j=1}^{B}}
\frac{(p_j^\perp(t)-P^{i\perp}_{env}(t))^{2}}{2m{B}}.
\end{equation}
$P^{i\perp}_{env}(t)$ is the mean transverse momentum of the B nucleons of 
the environment. The number of nucleons in the environment changes during 
the reaction. At the very end of the reaction there is usually no nucleon 
left in the environment because all fragments and single nucleons
are well separated in coordinate space.  $\Delta^{env}_{i}(t)$ can
sloppily be called "temperature of the
environment". If the fragments are only created at freeze out when the 
system is in thermal equilibrium 
we would expect that before freeze out there is no difference between 
the ``fragment temperature'' and the ``temperature of the environment''. 
This is due to the very fundamental fact that 
if a system is in thermal equilibrium all two or more body correlations 
are lost and therefore eventually existing correlations before freeze 
out cannot play a role for the production of the fragments at freeze out. 
In other words, every nucleon has the same chance to be finally part
of the fragment i and therefore the nucleons of the environment and 
the fragment nucleons should have the
same properties.

\begin{figure}[htb]
\epsfxsize=12.0cm
$$
\epsfbox{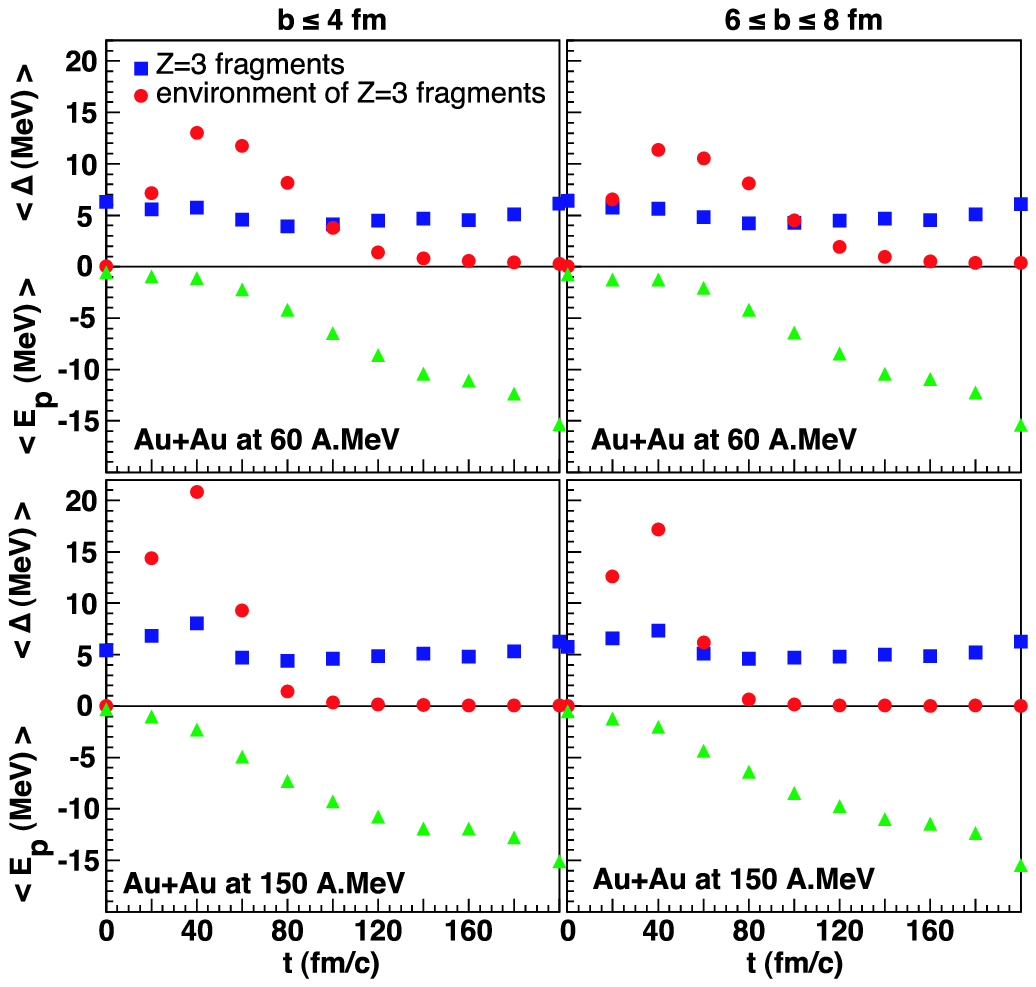}
$$
\caption{Time Evolution of the central density, of $\Delta $ and of
the binding energy Ep of Z=3 fragments during the reaction.}
\label{a12}
\end{figure}

The result of the simulations, averaged over all M fragments i of the 
size Z = 3, is displayed
in fig \ref{a12}. We see a quite different time evolution of $\Delta$
and $\Delta^{env}$. When passing the reaction zone, where the density is high, 
$\Delta^{env}$
increases strongly whereas $\Delta$ remains almost constant. The fragment 
nucleons do not take part in the heating of the system. 
At the end of the reaction 
$\Delta^{env}$ becomes smaller because the high relative momentum 
particles have already left the
environment. Only at the end of the reaction
$\Delta$ increases because the fragments leave the interaction zone with a 
deformed shape, as can be directly seen in the simulations. 
By regaining their spherical shape the (negative) binding 
energy increases and - due to energy conservation - the (positive) 
kinetic energy of nucleons in fragments
has to increase as well
in order to conserve the total energy.
At this late stage the fragments are separated, there are no environment 
nucleons around, therefore the "temperature of the environment" is 0. 
Thus we see that the mechanism we have found for the smaller Xe+Sn system \cite{goss}
is still 
valid  for the large Au+Au system. The mean free path of the Pauli blocked 
cross section for large transverse momentum transfer is still well below the
diameter of the system and therefore initial-final state correlations 
can survive.
\begin{figure}[htb]
\epsfxsize=12.0cm
$$
\epsfbox{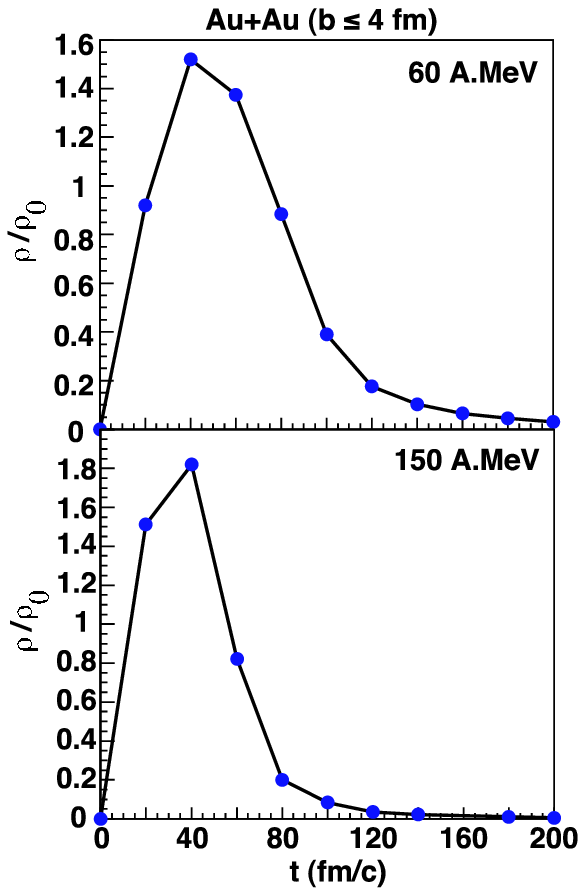}
$$
\end{figure}

\section{conclusion}
Using the QMD model we have investigated multifragmentation data in Au+Au 
reactions between 60 AMeV and 150 AMeV obtained by the INDRA collaboration. We observe
that in this energy range the transition between fragmentation of participant
matter and fragmentation of spectator matter takes place. We see for the first time
that midrapidity fragments are dominantly formed of an equal number of projectile
and target nucleons, as required if the system comes to equilibrium. This explains the 
success of statistical approaches in reproducing particle multiplicities in this phase space region.  
This mixing appears although the particle momentum does not equilibrate. 
Nucleons coming from the projectile (target) carry still a fraction of their initial 
collective momentum, especially if they end up in fragments. In fragments at midrapidity, which on average 
contain the same number of nucleons from projectile and target, these "memory effects"
compensate and do not influence the fragment momentum.
Free nucleons come closest to equilibrium because in order to be free they have usually suffered
collisions with a large momentum transfer.   
A common source of all fragments cannot be identified, the source
properties depend on the fragment size. This observation has been confirmed  
now independently from the data of the INDRA and the FOPI collaboration. 
Without passing through a filter theory and experiment cannot be compared. After 
filtering the QMD simulations describe well the energy dependent
event centrality, the multiplicity
distribution of fragments, the fragment yield and the distribution of the largest 
fragment. It describes as well the energy distribution of the smaller fragments.
Even for this large system, forward emitted fragments are initial state correlations   
in coordinate and/or momentum space which are not destroyed during the
reaction by collisions 
with a large  momentum transfer, similar to the observations for much smaller
systems. Thus for the largest system explored so far, the system comes close
to equilibrium but non-equilibrium effects still dominate outside a small midrapidity
zone.

\end{document}